\documentclass[journal]{IEEEtran}
\usepackage{tikz}

\usepackage{graphicx}
\usepackage{epsfig}
\usepackage{latexsym}
\usepackage{amsfonts}
\usepackage{here}
\usepackage{rawfonts}
\usepackage[latin1]{inputenc}
\usepackage[T1]{fontenc}
\usepackage{calc}
\usepackage{url}
\usepackage{enumerate}
\usepackage{color}
\usepackage[tbtags]{amsmath}
\usepackage{amssymb}
\usepackage{upref}
\usepackage{epic,eepic}
\usepackage{times}
\usepackage{dsfont}
\usepackage{comment}
\usepackage{cite}
\usepackage{algorithm}
\usepackage{algpseudocode}
\usepackage{graphicx}
\usepackage[font={small}]{caption}
\usepackage[font={small}]{subcaption}

\usepackage{stfloats}
\usepackage{mathtools}
\usepackage{amsmath}

\newtheorem{theorem}{\bf{Theorem}}
\newtheorem{definition}{\bf{Definition}}
\newtheorem{corollary}{\bf{Corollary}}

\newtheorem{remark}{Remark}

\ifCLASSINFOpdf
\else
\fi
\begin{document}

\title{Compute-Update Federated Learning: \\A Lattice Coding Approach Over-the-Air}

\author{{Seyed Mohammad Azimi-Abarghouyi} and Lav R. Varshney
  \thanks{S. M. Azimi-Abarghouyi is with the School of Electrical Engineering and Computer Science, KTH Royal Institute of Technology, Stockholm, Sweden (e-mail: seyaa@kth.se). L. R. Varshney is with the Department of Electrical and Computer Engineering, University of Illinois Urbana-Champaign, Urbana, IL USA (e-mail: varshney@illinois.edu). A part of this work was presented at IEEE ISIT 2024 \cite{conf}.} 
}

\maketitle

\begin{abstract}
This paper introduces a federated learning framework that enables over-the-air computation via digital communications, using a new joint source-channel coding scheme. Without relying on channel state information at devices, this scheme employs lattice codes to both quantize model parameters and exploit interference from the devices. We propose a novel receiver structure at the server, designed to reliably decode an integer combination of the quantized model parameters as a lattice point for the purpose of aggregation. We present a mathematical approach to derive
a convergence bound for the proposed scheme and offer
design remarks. In this context, we suggest an aggregation metric and a corresponding algorithm to determine effective integer coefficients for the aggregation in each communication round. Our results illustrate that, regardless of channel dynamics and data heterogeneity, our scheme consistently delivers superior learning accuracy across various parameters and markedly surpasses other over-the-air methodologies.

\end{abstract}

\section{Introduction}
In our current era, wireless edge devices like smartphones, autonomous vehicles, and sensors are becoming more advanced and ubiquitous. This evolution presents an opportunity to leverage machine learning techniques to build global models from the dispersed data these devices generate. Yet, streaming large datasets from these endpoints to centralized servers comes with numerous challenges, including data privacy, latency, power, and bandwidth limitations. Federated learning (FL) offers an innovative workaround to these obstacles \cite{mcmahan}. It allows devices to process machine learning locally, ensuring data stays on the device itself. In the FL paradigm, devices iteratively train models and send updates to a server, which aggregates them until convergence. This method proves invaluable in wireless settings, especially in IoT-rich ecosystems where network reliability and resource availability are often compromised. Given that these devices interact with the edge server over a shared wireless medium, FL also presents an intricate communication puzzle to solve. While FL can even be leveraged to enhance wireless communications  \cite{pooor}, traditional wireless communication techniques used in FL rely on orthogonal multiple access techniques. While these techniques prevent interference by requiring individual transmissions from each device to the server, they introduce significant communication latency and demand substantial resources \cite{viktoria}.

Harnessing interference from concurrent multi-access transmissions of edge devices, over-the-air computation \cite{nazer, power_huang, varsh} emerges as a promising strategy. Building on this foundation, a method termed over-the-air FL has been developed to handle the aggregation phase of FL, in the presence of interference, all within a single resource block \cite{viktoria}. This integration of communication and computation allows over-the-air FL to function more efficiently, consuming fewer resources and achieving lower latency than FL using orthogonal transmissions. However, previous research on over-the-air FL has assumed analog modulation, where the transmitter can freely shape the carrier waveform by opting for any real number I/Q coefficients \cite{huang_analog, ding, gunduz2, cao, ng, azimi_FL, schober, azimi_FL_conf, bereyhi, latif}. This assumption may not be possible for the majority of existing wireless devices, as they come with digital modulation chips that may not support arbitrary modulation schemes. 

Moreover, with a power control approach for channel compensation, past work required perfect channel state information at the transmitter (CSIT) for all devices to determine their transmission powers and phases to counteract wireless channel effects during the aggregation process. This approach requires devices to have extra hardware for accurate channel adjustments. Moreover, under poor channel conditions, a device might be unable to participate in the learning process or it could need high transmission power. This is problematic due to the physical constraints on both the instantaneous and average power capabilities of the device \cite{huang_analog,ding, gunduz2, gunduz3, cao, ng, azimi_FL, schober, azimi_FL_conf, bereyhi, latif}. One strategy for power control is truncated power control, which is suitable for scenarios with a single-antenna server \cite{huang_analog, schober, cao, azimi_FL, azimi_FL_conf,gunduz3}. In this case, each device only needs to know its own channel, known as local CSIT. Another strategy is joint device selection and power control schemes as suggested in \cite{ding, ng, latif, bereyhi}. These studies require global channel knowledge of all devices before each transmission, referred to as global CSIT, in order to centralize their power optimization process. In essence, the device selection strategy seeks to maximize the number of devices included in each communication round. This issue generally corresponds to an integer program, which is inherently NP-hard. Past work also requires perfect synchronization among the transmitters \cite{huang_analog, schober, azimi_FL, cao, azimi_FL_conf,gunduz3}. These requirements lead to a substantial overhead for channel estimation training and complex feedback mechanisms before any transmission, causing increased delays and reduced spectral efficiency.

These objectives contradict the aim of FL to enable low-cost distributed learning for a wide range of digital devices constrained by power and hardware limitations \cite{smith}. The special case of over-the-air FL using BPSK modulation is studied in \cite{gunduz3}. Nonetheless, while significantly reducing implementation cost and resource requirements, one-bit quantization can result in high delays and performance gap with the non-quantized case \cite{onebit}. Furthermore, other challenges related to channel estimation and power control continue to persist. 

Many wireless systems use constant power for blind transmission, not directly adjusting for the channel. Beyond not requiring CSIT, this approach can offer a range of advantages. Firstly, it enables the maintenance of average energy for signal transmission, no matter the variations in the channel. Additionally, it helps prevent enlarging the dynamic range of the signal being transmitted, making hardware implementations significantly simpler and reducing costs. Lastly, approaches that adjust for the channel at the transmitter can run into channel estimation errors, causing the values to be multiplied by unpredictable gains when received \cite{cohen1}. Considering these, the concept of a blind over-the-air FL approach has gained prominence. In \cite{gunduz4, turky, amiri1, amiri2}, there are strategies for blind over-the-air FL that rely on the channel state information at the receiver (CSIR). However, these methods call for a large number of receiver antennas, and the impact of wireless fading diminishes as the antenna count grows without bound. Additionally, the works in \cite{amiri1, amiri2} require sparsity conditions on gradient parameters. On the other hand, \cite{cohen1, cohen2, yang} investigate blind over-the-air FL that does not compensate for the adverse effects of fading. 

This work introduces an FL framework, which does not need CSIT, and extends over-the-air FL to scenarios where transmitters can use digital modulation. Our fresh perspective on over-the-air FL uses lattice codes with adjustable quantization levels to create an integrated end-to-end structure that is uniquely tailored to FL, encompassing quantization, transmission, and aggregation phases.

\subsection{Prior Work}
In \cite{eldar, eldar2}, a \textit{source coding scheme} based on lattice codes has been used to quantize model parameters in FL. In \cite{eldar2}, privacy enhancement is also considered. In these works, orthogonal transmission is used by the individual transmitters to prevent interference. In fact, after quantization, each transmitter maps quantized model parameters to bits and use conventional orthogonal digital communications. This use of lattice-based quantization is based on the principles of subtractive dithered lattice quantization \cite{rzamir1}, which is grounded in information-theoretic arguments. This method is recognized for its capability to achieve the highest possible finite-bit representation accuracy as per rate-distortion theory, with a margin of error that can be controlled \cite{rzamir2, eldar}. The proposed approach in \cite{eldar} achieves a more accurate quantized representation for FL applications compared to traditional scalar quantization techniques, including both probabilistic and deterministic approaches, e.g., \cite{huang_qnt}. In a sense, it uses functional quantization \cite{varsh2}.

Previous studies \cite{nazer_cmp, caire, azimi_cf1, azimi_cf2, azimi_cf3} have separately shown that lattice codes have benefits when used in compute-and-forward relaying as a \textit{channel coding scheme} \cite{nazer_cmp}. The lattice structure, where any integer combination of codewords is a codeword itself, allows for interference to be exploited during transmission and decoding, resulting in high bit-rate throughput. In the compute-and-forward framework, the assumption is that bits are already present at the transmitters, and nested lattice coding defined over a finite field is employed for modulation to provide end-to-end bit transmission. 
The main reason to use such integer combinations in compute-and-forward is because they can be decoded at much higher rates than individual messages at the relays, despite these combinations not having inherent explicit meaning. Based on the computation rate metric from information theory, multiple independent combinations with highest rates are chosen from different relays and finally transmitted to the receiver for decoding individual messages \cite{nazer_cmp, azimi_cf1}.

\subsection{Key Contributions}
We propose a \textit{joint source-channel coding scheme} that incorporates novel transmission and aggregation strategies, aiming to boost the resistance of over-the-air FL to interference and noise and achieve the desired learning outcomes. This scheme is accompanied by innovative digital transmitter and receiver architectures that leverage the lattice structure. Notably, it does not rely on any prior knowledge or CSIT, leading to a blind approach. Our scheme effectively mitigates errors resulting from learning and communication impairments, providing robustness in the presence of channel uncertainties and under a limited number of antennas at the server.

\textit{Compute-Update Scheme and its Transmitter-Receiver Architecture:} We provide an end-to-end real-valued model parameter transmission framework for FL, named compute-update FL--- {\fontfamily{lmtt}\selectfont
	FedCPU}. On the transmitter side, we use lattice codes over the infinite field of real numbers for quantization of the model parameters. Our quantization approach is based on normalization and dithering. On the server side, we decode a single integer combination of transmitted quantized model parameters as a lattice point, which after processing yields a new adjustable form of aggregation. 
Here, our scheme assigns a clear and explicit significance to integer combinations. This is due to the additive structure of model parameter aggregation in FL, where integer coefficients can be directly and purposefully linked to aggregation weights, giving them a distinct meaning and function. Additionally, lattice Voronoi region over the decoded integer combination protects against the decoding error arising from interference and noise. This level of protection is absent in analog modulation schemes, given their continuous transmitted values \cite{huang_analog,ding, gunduz2, cao, ng, azimi_FL, bereyhi, latif, cohen1, cohen2, yang, gunduz4, turky, schober, azimi_FL_conf, amiri1, amiri2}. Also, the aggregation method we propose marks a significant shift from the traditional practice in FL of employing fixed, predefined weights, which are either equal or based on the size of local datasets. This conventional approach, recommended for ideal, error-free communication conditions as outlined in the seminal FL paper \cite{mcmahan}, has been consistently used in over-the-air FL research \cite{huang_analog,ding, gunduz2, gunduz3, cao, ng, azimi_FL, bereyhi, latif, cohen1, cohen2, yang, gunduz4, turky, schober, azimi_FL_conf, amiri1, amiri2}. However, this persists despite the fact that over-the-air FL deals with imperfect communication scenarios, plagued by interference and noise. 

In response to this, our method of aggregation is designed to strategically tailor the aggregation weights. This is crucial because in our model, each integer combination results in a unique decoding error, leading to substantial variability. In order to accomplish the proposed aggregation, we propose a two-layer receiver architecture at the server side that incorporates both the real and imaginary parts of the signal. The initial layer incorporates an equalization vector, which is subsequently optimized to minimize the decoding error. In the subsequent layer, a normalizing factor is introduced and optimized to minimize the quantization error. This architecture is distinct from the single-layer receiver of the compute-and-forward \cite{nazer_cmp}, which also primarily relies on modulo operations.

\textit{Convergence Analysis and Aggregation Metric:} We characterize the convergence of {\fontfamily{lmtt}\selectfont
	FedCPU} in terms of the optimality gap and offer recommendations for its design. Based on the analysis, we present a metric for selecting integer coefficients for the proposed aggregation in each communication round. The metric depends on communication and learning factors, including quantization and decoding errors as well as a mismatch term between the integer and ideal coefficients. This justifies our joint communication and learning approach. We also suggest a convexification method and accordingly an efficient algorithm to optimize the metric. Our approach enables customizing the learning process over wireless networks to optimize the learning performance, as should be the goal for any learning task. In contrast, conventional methods such as those discussed in \cite{eldar, eldar2, huang_qnt} involve separate designs for quantization and transmission, with the main goal of maximizing transmission rate.

\textit{System Insights:} Our experimental results highlight the efficacy of {\fontfamily{lmtt}\selectfont
	FedCPU} in addressing the challenges posed by the lack of CSIT and the presence of only a limited number of antennas at the server. Enhancing the number of antennas bolsters performance. However, shrinking the lattice Voronoi regions reveals a balance between increased decoding error and diminished quantization error. Notably, {\fontfamily{lmtt}\selectfont
	FedCPU} showcases markedly superior learning accuracy compared to other over-the-air schemes, including the CSIT-equipped scheme \cite{ding}, due to its distinctive aggregation technique and finely-tuned receiver design. Furthermore, the learning outcomes are very close to what one might expect in an ideal situation with orthogonal transmission, even when the number of antennas is limited. Additionally, the choice of lattice code used impacts the results.

\section{System Model}
\begin{figure}[tb!]
	\vspace{-20pt}
\hspace{-10pt}
\includegraphics[width =3.5in]{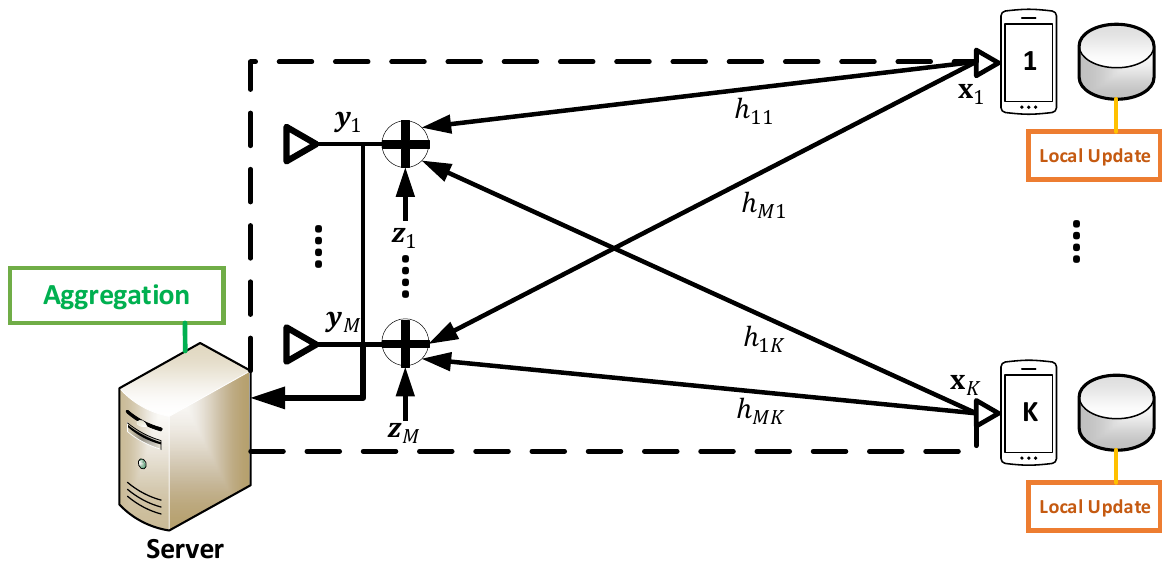}\hspace{-15pt}
\vspace{-200pt}
\caption{Compute-update FL system.}
\vspace{-5pt}
\end{figure}
\subsection{Setup}
There are $K$ devices and a single server as the basic setup for FL systems, see Fig. 1. All the devices are single-antenna units, but the server has $M$ antennas. There is no prior knowledge of local data statistics of the devices, consistent with \cite{huang_analog,ding, gunduz2, gunduz3, cao, ng, azimi_FL, bereyhi, latif, cohen1, cohen2, yang, gunduz4, turky, schober, azimi_FL_conf, amiri1, amiri2}. The downlink channels from the server to the devices are considered error-free. The validity of this assumption comes from the server's high transmission power, its efficient antenna capability,
and the exclusive transmission to the devices in the downlink. The uplink channel from each device $k$ to the $m$-th antenna of the server at communication round $t$ is represented by ${h}_{mk,t} = |{h}_{mk,t}| e^{\angle {h}_{mk,t}} \in {\mathbb{C}}$, where $|{h}_{mk,t}|$ is the channel gain and $\angle {h}_{mk,t}$ is the channel phase\footnote{Partial asynchrony among the devices can be modeled as part of the channel phase.}. The $\angle {h}_{mk,t}$ spans the entire range from $0$ to $2\pi$. All channels are assumed invariant during one time slot required for an uplink transmission, while they change independently from one time slot to another. Furthermore, all devices have frame-level synchronization with no carrier frequency offset, and are considered to have the same transmission power constraint $P$. However, by appropriately adjusting the channel coefficients, we can integrate asymmetric power constraints. Let the entire channel matrix $\mathbf{H}_{t}^{\text{c}} \in {\mathbb{C}}^{M \times K}$ be 
\begin{align}
\mathbf{H}_{t}^{\text{c}} =
\begin{bmatrix}
{h}_{11,t} & \cdots & {h}_{1K,t} \\
\vdots & \ddots & \vdots \\
{h}_{M1,t} & \cdots & {h}_{MK,t} \\
\end{bmatrix}.
\end{align}
This channel model is extensively adopted in several other over-the-air FL schemes \cite{huang_analog,ding, gunduz2, gunduz3, cao, ng, azimi_FL, bereyhi, latif, cohen1, cohen2, yang, gunduz4, turky, schober, azimi_FL_conf, amiri1, amiri2}. The server is the only node that knows $\mathbf{H}_t^{\text{c}}$ as CSIR \textit{after the transmission}, whereas the devices have no any knowledge of channels. Accurate CSIR is acquired through the individual uplink transmission of training sequences from each device to the server.\footnote{Consistent with previous works \cite{huang_analog, ding, gunduz2, cao, ng, azimi_FL, bereyhi, latif, schober, azimi_FL_conf, amiri1, amiri2}, we do not account for channel estimation error in this study.} This is the main assumption in the blind over-the-air FL strategy \cite{turky, gunduz4,amiri1, amiri2} in line with many wireless systems that use transmissions with constant power, avoiding channel compensation at the transmitter side \cite{nazer_cmp, caire, azimi_cf1, azimi_cf2, azimi_cf3, azimi_cf4, ordi,nazermimo}. Such a strategy maintains a steady average signal energy regardless of channel variations, limits the signal's dynamic range, and reduces device hardware complexity. It emerges as a particularly promising solution for IoT applications equipped with low-capability devices. Nevertheless, the majority of over-the-air FL methods with power control strategy, e.g., \cite{ding, ng, bereyhi, latif}, need both CSIR and CSIT \textit{before each transmission}.

\subsection{Learning Algorithm}
Device $k \in \{1,\ldots,K\}$ has its own local dataset $\mathcal{D}_k$. The learning model is characterized by the parameter vector $\mathbf{w} \in \mathbb{R}^{s\times 1}$, where $s$ indicates the model's size. Subsequently, the local loss function for the model vector $\mathbf{w}$ when applied to $\mathcal{D}_k$ is
\begin{align}
F_k(\mathbf{w}) =  \frac{1}{|\mathcal{D}_k|}\sum_{\xi_i\in {\cal D}_k}^{}f(\mathbf{w},\xi_i),
\end{align}
where $f(\mathbf{w},\xi_i)$ represents the per-sample loss function, capturing the prediction error of $\mathbf{w}$ for a given sample $\xi_i$. As a result, the cumulative loss function across all the distributed datasets $\bigcup_{k=1}^{K} \mathcal{D}_k$ can be expressed as
\begin{align}
\label{lossfunction}
F(\mathbf{w}) = \frac{1}{\sum_{k=1}^{K}|\mathcal{D}_k|}\sum_{k=1}^{K} |\mathcal{D}_k| F_k(\mathbf{w}).
\end{align}

The goal of the training procedure is to identify an optimal parameter vector $\mathbf{w}$ that minimizes $F(\mathbf{w})$. This can be formulated as
\begin{align}
\label{objective}
\mathbf{w}^* = \min_{\mathbf{w}} F(\mathbf{w}).
\end{align}

Conventional FL algorithms address the problem posed in \eqref{objective} through a two-step process: local update and subsequent aggregation, cycling through several rounds. Within this structure, a frequently adopted FL algorithm named {\fontfamily{lmtt}\selectfont FedAvg} \cite{mcmahan} is elucidated as follows.

For a specific round $t$ within the set $\{0,\ldots,T-1\}$, where $T$ is the total number of rounds, every device $k$ initiates by refining its individual learning model over $\tau$ local stochastic gradient
descent steps, each based on a mini-batch $\boldsymbol\xi_k^i$ of size $B$ selected at random from $\mathcal{D}_k$. This process is further detailed as
\begin{align}
\mathbf{w}_{k,t,i+1} = {\mathbf{w}_{k,t,i}}- \mu_t \nabla F_k(\mathbf{w}_{k,t,i}, \boldsymbol\xi_k^i), \forall i \in \left\{0,\ldots,\tau-1\right\},
\end{align}
where $\mu_t$ is the learning rate at round $t$. Then, each device $k$ uploads the local model update $\Delta\mathbf{w}_{k,t} = {\mathbf{w}_{k,t,\tau}} - {\mathbf{w}_{k,t,0}}$ to the server for aggregation. For ideal aggregation, the global model update can be achieved by averaging the model updates from all devices, giving them equal weights, as
\begin{align}
\label{agg}
\Delta\mathbf{w}_{\text{G},t+1} = \mathbf{w}_{\text{G},t+1} - \mathbf{w}_{\text{G},t} = \frac{1}{K}\sum_{k=1}^{K} \Delta\mathbf{w}_{k,t}.
\end{align}
Subsequently, the server broadcasts the acquired global model $\mathbf{w}_{\text{G},t+1}$ to all devices. Using this model, each device $k$ sets its starting state for the upcoming round as $\mathbf{w}_{k,t+1,0} = \mathbf{w}_{\text{G},t+1}$. This cycle continues until the predetermined number of rounds $T$ is reached.

\subsection{Lattice Preliminaries}
\begin{definition}
A lattice $\Lambda$ is a discrete subgroup of $\mathbb{R}^{s \times 1}$ and can be expressed as a linear transformation of integer vectors as
\begin{align}
\Lambda = \left\{\mathbf{G}\mathbf{s}: \mathbf{s} \in \mathbb{Z}^{s \times 1}\right\},
\end{align}
where $\mathbf{G} \in \mathbb{R}^{s\times s}$ is the lattice generator matrix. 

The lattice $\Lambda$ has two main features: For any points $\mathbf{x}_1, \mathbf{x}_2 \in \Lambda$, we have $\mathbf{x}_1+\mathbf{x}_2 \in \Lambda$ and for any $\mathbf{x} \in \Lambda$, we have $a\mathbf{x}_1 \in \Lambda, \forall a \in \mathbb{Z}$. Consequently, any linear integer combination of lattice points is itself a lattice point.
\end{definition}

Lattice codebooks can be formed
by combining a regular low-dimensional lattice constellation, such as pulse amplitude modulation (PAM) or quadrature amplitude modulation (QAM), with a linear code like LDPC \cite{ordi, nazermimo}. This integration enables the feasible implementation of lattice codes in common digital modulators. Methods
for designing the lattice generator matrix $\mathbf{G}$ can be also found
in \cite{agrell}.

\begin{definition}
A lattice quantizer is a map $Q_{\Lambda}: \mathbb{R}^{s \times 1} \to \Lambda$ that quantizes a point to its nearest lattice point based on
Euclidean distance as
\begin{align}
Q_{\Lambda} (\mathbf{x}) = \arg\min_{\lambda \in \Lambda} \|\mathbf{x}-\lambda\|^2. 
\end{align}
\end{definition}

\begin{definition}
Fundamental Voronoi region $\mathcal{V}$ is the set of all
points that are quantized to the origin as
\begin{align}
\mathcal{V} = \left\{\mathbf{x} \in \mathbb{R}^{s \times 1} : Q_{\Lambda}(\mathbf{x}) = \mathbf{0}\right\}.
\end{align}
\end{definition}

\begin{definition}
The second moment of a lattice is defined as the per-dimension second moment of a uniform distribution over the foundational Voronoi region $\cal V$ as
\begin{align}
\sigma_\text{q}^2 = \frac{\int_{\mathcal{V}}^{}\|\mathbf{x}\|^2\mathrm{d}\mathbf{x}}{s\int_{\mathcal{V}}^{}\mathrm{d}\mathbf{x}}.
\end{align}

\end{definition}

\section{FedCPU: Compute-Update Scheme}
The {\fontfamily{lmtt}\selectfont
	FedCPU} consists of two primary components: the transmission scheme employed by the devices and the aggregation scheme utilized by the server. Through simultaneous transmission, {\fontfamily{lmtt}\selectfont FedCPU} introduces a novel adjustable version of the aggregation vector, where model updates from devices are weighted by integer coefficients that are not predefined. This is because each set of integer coefficients results in a unique learning performance. This version exploits the lattice structure and the additive nature of wireless multiple-access channels, given as
\begin{align}
\label{w_agg}
\Delta\mathbf{w}_{\text{G},t+1}  = \frac{1}{\mathbf{1}^\top \mathbf{a}_t}\sum_{k=1}^{K} a_{k,t} \Delta\mathbf{w}_{k,t},
\end{align}
where $a_{k,t}$ is the integer coefficient corresponding to device $k$, and the integer coefficient vector $\mathbf{a}_t = [a_{1,t},\ldots,a_{K,t}]^\top \in \mathbb{Z}^{K \times 1}$. The summation $\mathbf{1}^\top\mathbf{a}_t$ is to take the average for the aggregation. One can consider any vector $\mathbf{a}$ with non-negative integer coefficients except the all-zero vector $\mathbf{0}$.  {\fontfamily{lmtt}\selectfont FedCPU} operates with constant power and does not rely on prior information. This sets {\fontfamily{lmtt}\selectfont FedCPU} apart from over-the-air FL schemes that use power control, imposing average and maximum power constraints, and necessitate device selection \cite{huang_analog,ding, gunduz2, gunduz3, cao, ng, azimi_FL, schober, azimi_FL_conf, bereyhi, latif}. With {\fontfamily{lmtt}\selectfont FedCPU}, every device participates in the learning process based on an appropriate aggregation weight tailored to its communication conditions, free from the mentioned constraints. Additionally, while Subsection II.B details a popular learning algorithm, {\fontfamily{lmtt}\selectfont FedCPU} is not confined to it. For the sake of presentation simplicity, we have omitted the iteration index in this section.
\subsection{Transmission Scheme} Based on a lattice $\Lambda$, the transmission preparation procedure carried out by each device $k$ is as follows.

1) The model update parameters undergo normalization to achieve zero mean and unit variance. This normalization serves two purposes. Firstly, zero-mean entries ensure the subsequent estimate of aggregation remain unbiased. Secondly, unit variance ensures that the power of the received signal remains independent of the particular values of the model update parameters. 

The normalized model update is obtained as ${\widehat{\Delta\mathbf{w}}_{k}} = \frac{{\Delta{\mathbf{w}}_k}-\vartheta_k\mathbf{1}}{\sigma_k}$, where $\mathbf{1}$ is the all-one vector, and $\vartheta_k$ and ${\sigma_k}$ denote the mean and standard deviation of the $s$ entries of the model update vector given by
\begin{align}
\vartheta_k = \frac{1}{s}\sum_{i=1}^{s} \delta w_{k,i},\
\sigma_k^2 = \frac{1}{s}\sum_{i=1}^{s}(\delta w_{k,i}-\vartheta_k)^2,
\end{align}
where $\delta w_{k,i}$ is the $i$-th entry of the vector $\Delta \mathbf{w}_k$. Each device $k$ shares the two scalars $(\vartheta_k,\sigma_k)$ error-free to the server via an orthogonal feedback channel.

2) A dither vector $\mathbf{d}_{k} \in \mathcal{V}$
is generated independently from other devices according to a random uniform distribution.
 
3) The dither is added to $\widehat{\Delta\mathbf{w}}_k$ and the nearest codeword to the result is selected as
\begin{align}
\label{qnt_fun}
{\overline {\Delta\mathbf{w}}_{k}} = Q_{\Lambda}( \widehat {\Delta\mathbf{w}}_k+\mathbf{d}_{k}).
\end{align}
The addition of dither results in the quantization error becoming uniform, allowing us to statistically describe this error using the second moment of the lattice. For the quantization error $\boldsymbol\epsilon_k$ defined as in $Q_{\Lambda}( \widehat {\Delta\mathbf{w}}_k+\mathbf{d}_{k})-\mathbf{d}_k = \widehat{\Delta \mathbf{w}_k} + \boldsymbol\epsilon_k$, it holds that $\mathbb{E}\left\{\|\boldsymbol\epsilon_k\|^2\right\} = \sigma_\text{q}^2$. This dither also aids in ensuring the transmitted points from various devices are uncorrelated. Also, notice that the transmit normalization limits the number of possible points resulting from \eqref{qnt_fun}.

4) The device scales the resulting lattice point and transmits
\begin{align}
\mathbf{x}_{k} =  \sqrt{\frac{P}{1+2\sigma_\text{q}^2}} {\overline {\Delta\mathbf{w}}_{k}},
\end{align}
whereby we have
\begin{align} 
&\mathbb{E}\left\{\|\mathbf{x}_k\|^2\right\} = \frac{P}{1+2\sigma_\text{q}^2}\mathbb{E}\left\{\|{\overline {\Delta\mathbf{w}}_{k}}\|^2\right\} \leq \frac{P}{1+2\sigma_\text{q}^2}\nonumber\\&\times \left(\mathbb{E}\left\{\|\widehat {\Delta\mathbf{w}}_k+\mathbf{d}_{k}\|^2\right\} + \sigma_\text{q}^2\right)= \frac{P}{1+2\sigma_\text{q}^2}\nonumber\\&\times \left(\mathbb{E}\left\{\|\widehat {\Delta\mathbf{w}}_k\|^2\right\}+\mathbb{E}\left\{\|\mathbf{d}_{k}\|^2\right\} + \sigma_\text{q}^2\right) = P,
\end{align}
which satisfies the power constraint at the devices.

The server has knowledge of all the dither vectors, from a shared common randomness between the server and the devices. Such assumption is widely adopted in the literature \cite{eldar, eldar2,rzamir1,rzamir2,nazer_cmp, caire, azimi_cf1, azimi_cf2, azimi_cf3,  azimi_cf4, ordi,nazermimo}. Also, notice that none of the above operations depend on the channels. Fig. 2
provides a schematic representation of the transmitter architecture.
\begin{figure}[tb!]
	\centering
	
	\includegraphics[width =3.6in]{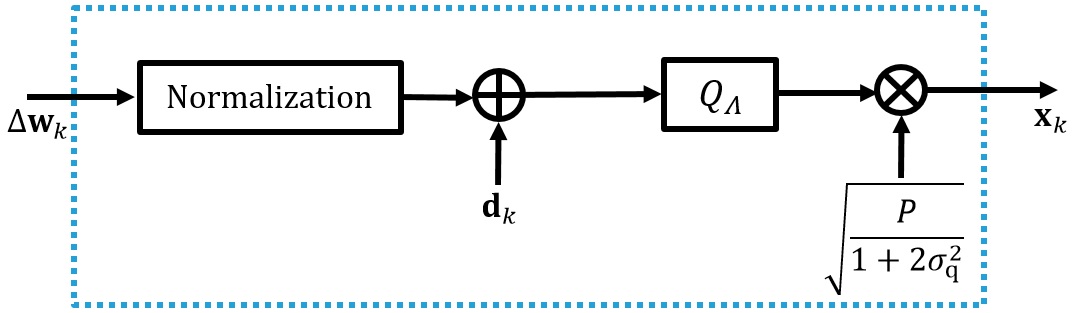}
	
	\caption{Transmitter structure for device $k$.}
	\vspace{0pt}
\end{figure}
\subsection{Aggregation Scheme} 
After simultaneous transmission by all the devices in a single resource block, the baseband received signal at antenna $m$ of the server, $\mathbf{y}_m \in \mathbb{C}^{{s}\times 1}$, is
\begin{align}
\label{real_signal}
\mathbf{y}_m = \sum_{k=1}^{K}h_{mk} \mathbf{x}_k + \mathbf{z}_m,
\end{align} 
where $\mathbf{z}_m \in \mathbb{C}^{{s}\times 1}$ is complex Gaussian noise, with each entry having a variance of $\sigma_\text{z}^2$. 
It is the standard multiple access communication model \cite{huang_analog,ding, gunduz2, gunduz3, cao, ng, azimi_FL, cohen1, cohen2, yang, gunduz4, turky, schober, azimi_FL_conf, amiri1, amiri2, nazer_cmp, caire, azimi_cf1, azimi_cf2, azimi_cf3, azimi_cf4, ordi, bereyhi, latif}. Thus, the real-valued representation of \eqref{real_signal} is as follows.
\begin{align}
\label{realsignal}
\mathbf{Y} = {\mathbf{H}}\mathbf{X} + \mathbf{Z},
\end{align}
where
\begin{align}
\mathbf{Y} = \begin{bmatrix}
\mathfrak{Re}\left\{\mathbf{y}_1^\top\right\} \\ \vdots \\ \mathfrak{Re}\left\{\mathbf{y}_M^\top\right\}\\
\mathfrak{Im}\left\{\mathbf{y}_1^\top\right\}\\ \vdots \\ \mathfrak{Im}\left\{\mathbf{y}_M^\top\right\}
\end{bmatrix} \in \mathbb{R}^{2M \times s},\ \mathbf{Z} = \begin{bmatrix}
\mathfrak{Re}\left\{\mathbf{z}_1^\top\right\} \\ \vdots \\ \mathfrak{Re}\left\{\mathbf{z}_M^\top\right\}\\
\mathfrak{Im}\left\{\mathbf{z}_1^\top\right\}\\ \vdots \\ \mathfrak{Im}\left\{\mathbf{z}_M^\top\right\}
\end{bmatrix}\in \mathbb{R}^{2M \times s}.
\end{align}
\vspace{-20pt}
\begin{align}
{\mathbf{H}} = \begin{bmatrix}
\mathfrak{Re}\left\{\mathbf{H}^\text{c}\right\}  \\
\mathfrak{Im}\left\{\mathbf{H}^\text{c}\right\}
\end{bmatrix}\in \mathbb{R}^{2M \times K},\ \mathbf{X} = \begin{bmatrix}
\mathbf{x}_1^\top \\ \vdots \\ \mathbf{x}_K^\top
\end{bmatrix}\in \mathbb{R}^{K \times s},
\end{align}
Thus, the resource efficiency, in terms of the number of communication resources used per any $1 \leq K <\infty$ devices, is $\frac{1}{K}$. Based on \eqref{realsignal} and the transmission scheme in Subsection III.A, the receiver architecture for the aggregation at the server includes two layers as follows.
\subsubsection{First Layer}

An equalization vector $\mathbf{b} \in \mathbb{R}^{2 M \times 1}$ is used to obtain
\begin{align}
\mathbf{b}^\top \mathbf{Y} = \mathbf{b}^\top {\mathbf{H}} \mathbf{X}+ \mathbf{b}^\top \mathbf{Z}.
\end{align}
Then, the decoder aims to recover a lattice point or equally a linear integer combination of quantized model updates $\sum_{k=1}^{K}a_k \overline{\Delta\mathbf{w}}_k^\top = \mathbf{a}^\top \overline{\Delta\mathbf{W}}$ directly from $\mathbf{b}^\top \mathbf{Y}$, and
\begin{align}
\overline{\Delta\mathbf{W}} = \begin{bmatrix}
\overline{\Delta\mathbf{w}}_{1}^\top \\
\vdots \\
\overline{\Delta\mathbf{w}}_{K}^\top 
\end{bmatrix}\in \mathbb{R}^{K \times s},
\end{align}
For this decoding, the result $\mathbf{b}^\top \mathbf{Y}$ is scaled and then quantized to its nearest lattice point as
\begin{align}
Q_{\Lambda}\left(\sqrt{\frac{1+2\sigma_\text{q}^2}{P}}\mathbf{b}^\top \mathbf{Y}\right) = \mathbf{a}^\top \overline{\Delta\mathbf{W}}.
\end{align}

\begin{figure}[tb!]
	\centering
	
	\includegraphics[width =3.6in]{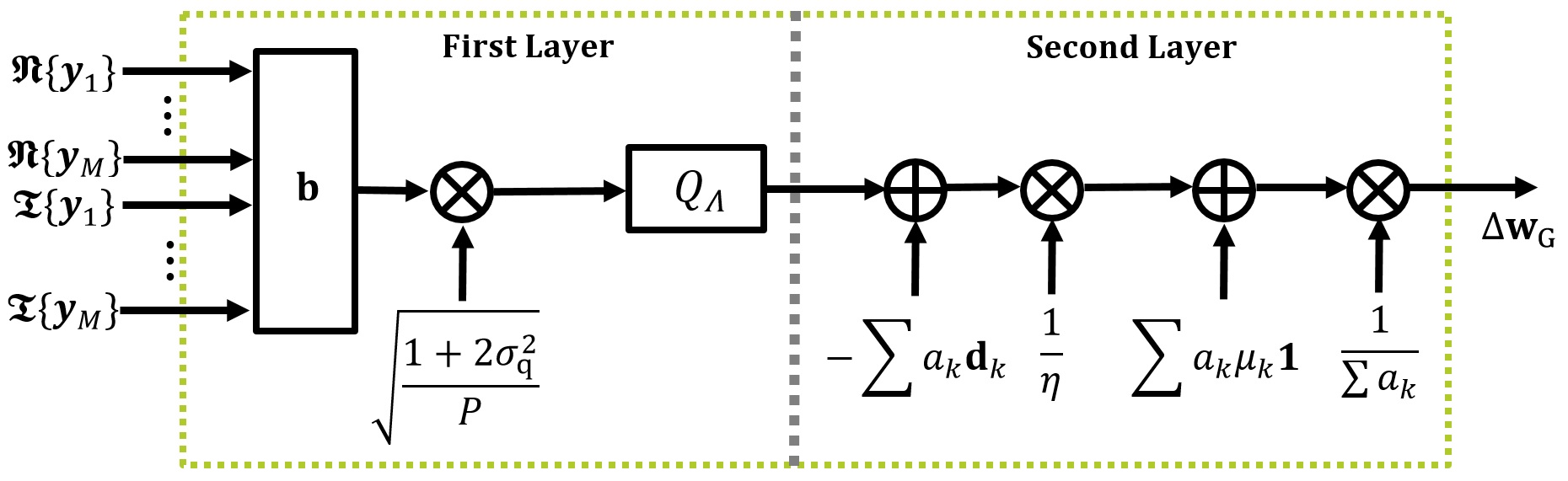}
	
	\caption{Receiver structure for the server.}
	\vspace{0pt}
\end{figure}
We can rewrite $\sqrt{\frac{1+2\sigma_\text{q}^2}{P}}\mathbf{b}^\top \mathbf{Y}$ as 
\begin{align}
\mathbf{a}^\top \overline{\Delta\mathbf{W}}+\sqrt{\frac{1+2\sigma_\text{q}^2}{P}}(\mathbf{b}^\top{\mathbf{H}}-\mathbf{a}^\top)\mathbf{X}+\sqrt{\frac{1+2\sigma_\text{q}^2}{P}}\mathbf{b}^\top \mathbf{Z}.
\end{align}
Thus, the decoding error in recovering $\mathbf{a}^\top \overline{\Delta\mathbf{W}}$ is measured in terms of the mean squared error (MSE), referred to as decoding MSE, as follows.
\begin{align}
\label{mse}
&\text{DMSE}(\mathbf{a}) = \mathbb{E}\Biggl\{\Biggl\Vert \sqrt{\frac{1+2\sigma_\text{q}^2}{P}}(\mathbf{b}^\top{\mathbf{H}}-\mathbf{a}^\top)\mathbf{X}+\sqrt{\frac{1+2\sigma_\text{q}^2}{P}}\times\nonumber\\
&\mathbf{b}^\top \mathbf{Z}\Biggr\Vert^2\Biggr\} =s({1+2\sigma_\text{q}^2})\left(\|\mathbf{b}^\top{\mathbf{H}}-\mathbf{a}^\top\|^2+\frac{1}{\text{SNR}}\|\mathbf{b}\|^2\right),
\end{align}
where the independence of $\mathbf{x}_k$ and $\mathbf{x}_{k'}$, $\forall k\neq k'$, is assumed. This independence assumption is supported by our proposed transmit normalization, the use of random independent dithers, and the execution of local mini-batch stochastic computations across different devices. Other over-the-air FL works \cite{power_huang, ding, ng, cao, azimi_FL, azimi_FL_conf, gunduz4, bereyhi, latif} also assume independence among transmitted signals. Also, $\text{SNR} = \frac{P}{\sigma_\text{z}^2}$ denotes the signal-to-noise ratio. The subsequent theorem presents the vector $\mathbf{b}$ that minimizes the decoding MSE.  
\begin{theorem}
The optimal equalization vector for a given coefficient vector $\mathbf{a}$ is
\begin{align}
\label{bopt}
\mathbf{b}_\text{opt}^\top = \mathbf{a}^\top {\mathbf{H}}^\top \left(\frac{1}{\text{SNR}}\mathbf{I}+{\mathbf{H}}{\mathbf{H}}^\top\right)^{-1}.
\end{align}
\end{theorem}
\begin{IEEEproof}
Expanding $\frac{\text{DMSE}(\mathbf{a})}{s({1+2\sigma_\text{q}^2})}$ as
\begin{align}
&\|\mathbf{b}^\top{\mathbf{H}}-\mathbf{a}^\top\|^2+\frac{1}{\text{SNR}}\|\mathbf{b}\|^2 = \left(\mathbf{b}^\top{\mathbf{H}}-\mathbf{a}^\top\right)\left({\mathbf{H}}^\top\mathbf{b}-\mathbf{a}\right)+\nonumber\\&\frac{1}{\text{SNR}}\|\mathbf{b}\|^2 = \mathbf{b}^\top{\mathbf{H}}{\mathbf{H}}^\top\mathbf{b} -2\mathbf{b}^\top{\mathbf{H}}\mathbf{a}+\mathbf{a}^\top\mathbf{a}+\frac{1}{\text{SNR}}\mathbf{b}^\top\mathbf{b},
\end{align}
and taking derivative from the result with respect to $\mathbf{b}$, we obtain
\begin{align}
2{\mathbf{H}}{\mathbf{H}}^\top \mathbf{b} - 2 {\mathbf{H}}\mathbf{a}+\frac{2}{\text{SNR}}\mathbf{b},
\end{align}
which amounts to zero at \eqref{bopt}.
\end{IEEEproof}
Substituting $\mathbf{b}_\text{opt}$ into \eqref{mse}, we obtain
\begin{align}
\label{dmse_mil}
&\text{DMSE}(\mathbf{a}) = s({1+2\sigma_\text{q}^2})\times \nonumber\\&\mathbf{a}^\top\left[ \mathbf{I} - {\mathbf{H}}^\top \left(\frac{1}{\text{SNR}}\mathbf{I}+{\mathbf{H}}{\mathbf{H}}^\top\right)^{-1}{\mathbf{H}}\right]\mathbf{a}.
\end{align}
Applying the matrix inversion lemma \cite{matrix_inversion}, \eqref{dmse_mil} can be expressed as
\begin{align}
\label{dmsee}
\text{DMSE}(\mathbf{a}) = s({1+2\sigma_\text{q}^2})\mathbf{a}^\top\left( \mathbf{I} +{\text{SNR}}{\mathbf{H}}^\top{\mathbf{H}}\right)^{-1}\mathbf{a}.
\end{align}
From the $\text{SNR}$ and $\mathbf{H}$ as seen in \eqref{dmsee}, the DMSE arises due to interference and noise impacting the decoding process. It is also worth noting that the matrix $\left( \mathbf{I} +{\text{SNR}}\mathbf{H}^\top\mathbf{H}\right)^{-1}$ is positive definite. 

The value of $\text{DMSE}(\mathbf{a})$ can be interpreted as the variance of the effective noise in decoding $\mathbf{a}^\top \overline{\Delta\mathbf{W}}$. Therefore, the decoding error probability, denoted as $P_{\text{D}}$, which represents the probability of decoding any other lattice point besides $\mathbf{a}^\top \overline{\Delta\mathbf{W}}$, is equated to the event where the decoding noise lies outside $\mathbf{a}^\top \overline{\Delta\mathbf{W}}+\mathcal{V}$.
\begin{remark}
For a decoding noise within $\mathbf{a}^\top \overline{\Delta\mathbf{W}}+\mathcal{V}$, the system is fully protected from interference and noise.
\end{remark}

\begin{remark}
Unlike conventional digital communications, where a decoding error results in an entirely incorrect message, in {\fontfamily{lmtt}\selectfont
	FedCPU} a decoding error merely introduces an additive estimation error in \eqref{decode_error_add}. This estimation error depends on the distance between the decoded lattice point and the target lattice point. In essence, while our goal is to minimize the decoding error to enhance system performance, it does not constitute a critical constraint or a bottleneck for the system. 
\end{remark}

\subsubsection{Second Layer} After decoding and removing the dithers, the proposed aggregation vector in \eqref{w_agg} at the output of the receiver can be estimated as
\begin{align}
\label{model_estimate}
\Delta\mathbf{w}_\text{G}^\top = \frac{Q_{\Lambda}\left(\sqrt{\frac{1+2\sigma_\text{q}^2}{P}}\mathbf{b}^\top \mathbf{Y}\right) - \mathbf{a}^\top {\mathbf{D}}}{\eta \mathbf{1}^\top\mathbf{a}}+\frac{1}{ \mathbf{1}^\top\mathbf{a}}\sum_{k=1}^{K} a_k\vartheta_k \mathbf{1}^\top,
\end{align}
where $\eta$ is a normalizing factor, and
\begin{align}
\mathbf{D} = \begin{bmatrix}
{\mathbf{d}}_{1}^\top \\
\vdots \\
{\mathbf{d}}_{K}^\top
\end{bmatrix} \in \mathbb{R}^{K \times s}.
\end{align}
Owing to the proposed transmit normalization, the mean of the initial term in \eqref{model_estimate} is zero, and the deliberate inclusion of the second term in \eqref{model_estimate} guarantees the precise mean recovery of \eqref{w_agg}. Consequently, the estimation in \eqref{model_estimate} is unbiased. It differes from \cite{ding, mcmahan, gunduz2, gunduz3, cao, ng, azimi_FL, bereyhi, latif, smith, cohen1, cohen2, yang, gunduz4, onebit, eldar, eldar2, turky, amiri1,amiri2, azimi_FL_conf, schober, huang_analog, huang_qnt} which use predefined weights in FL, such as equal or proportional to local dataset sizes. In Section IV, we quantify the effects of the proposed estimation given by \eqref{model_estimate} with an arbitrary vector $\mathbf{a}$ on learning convergence.
\begin{figure}[tb!]
		\vspace{-95pt}
	\hspace{-20pt}\includegraphics[width =4.3in]{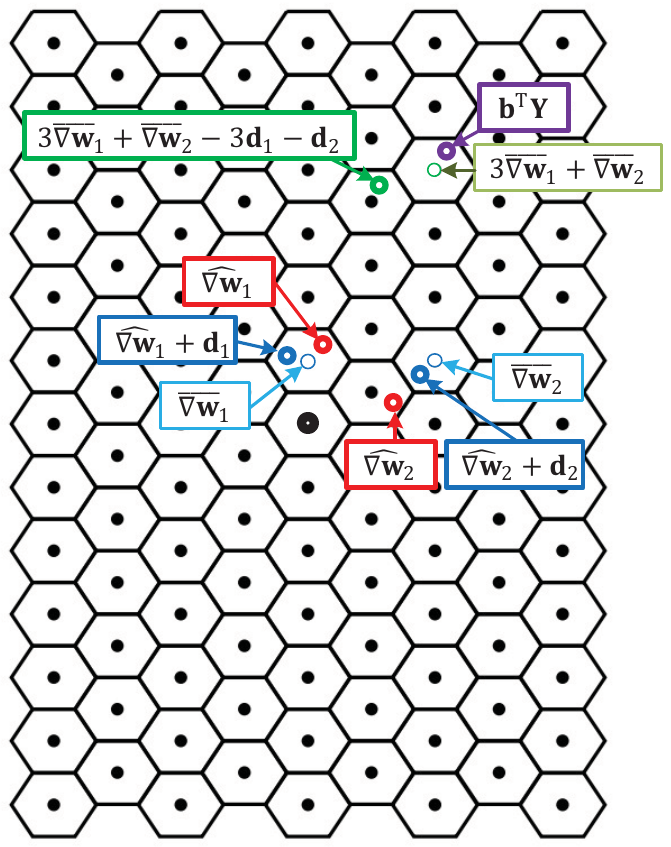}
	\vspace{-85pt}	
	\caption{Two-dimensional hexagonal lattice example, with a scenario including two devices. The shown green point $3 \overline{\Delta\mathbf{w}}_{1}+\overline{\Delta\mathbf{w}}_2- 3\mathbf{d}_1 - \mathbf{d}_2$ is processed for aggregation in the form of $\Delta\mathbf{w}_\text{G} = \frac{3 \overline{\Delta\mathbf{w}}_{1}+\overline{\Delta\mathbf{w}}_2- 3\mathbf{d}_1 - \mathbf{d}_2}{4\eta}+\frac{3 \vartheta_1 + \vartheta_2}{4}\mathbf{1}$, as in \eqref{model_estimate}.}
	\vspace{-10pt}
\end{figure}

We now prove that the aggregation expressed in \eqref{w_agg} can be derived from the estimation outlined in \eqref{model_estimate}. To illustrate this, we reformulate \eqref{model_estimate} as follows.
\begin{align}
\label{decode_error_add}
&\Delta\mathbf{w}_\text{G}^\top = \frac{1}{\mathbf{1}^\top \mathbf{a}} \sum_{k=1}^{K}a_k \Delta \mathbf{w}_k^\top + \frac{\mathbf{a}^\top\overline{\Delta\mathbf{W}}-\mathbf{a}^\top \mathbf{D}}{\eta \mathbf{1}^\top \mathbf{a}}+\nonumber\\
&\frac{1}{ \mathbf{1}^\top\mathbf{a}}\sum_{k=1}^{K} a_k \left(\vartheta_k\mathbf{1}^\top - \Delta \mathbf{w}_k^\top\right) =\frac{1}{\mathbf{1}^\top \mathbf{a}} \sum_{k=1}^{K}a_k \Delta \mathbf{w}_k^\top +\nonumber\\
& \frac{\mathbf{a}^\top\overline{\Delta\mathbf{W}}-\mathbf{a}^\top \mathbf{D}}{\eta \mathbf{1}^\top \mathbf{a}}-\frac{1}{\mathbf{1}^\top \mathbf{a}}\sum_{k=1}^{K}a_k\sigma_k\widehat{\Delta \mathbf{w}_k}^\top =\frac{1}{\mathbf{1}^\top \mathbf{a}} \sum_{k=1}^{K}a_k \Delta \mathbf{w}_k^\top \nonumber\\
&+ \frac{\sum_{k=1}^{K}a_k\widehat{\Delta\mathbf{w}_k}^\top+\sum_{k=1}^{K}a_k \boldsymbol\epsilon_{k}^\top}{\eta \mathbf{1}^\top \mathbf{a}}-\frac{1}{\mathbf{1}^\top \mathbf{a}}\sum_{k=1}^{K}a_k\sigma_k\widehat{\Delta \mathbf{w}_k}^\top.
\end{align}
In \eqref{decode_error_add}, the first term is the aggregation \eqref{w_agg} and the remaining terms are mainly due to the quantization errors. Thus, the overall error against this estimation in terms of the MSE, referred to as qunatization MSE, is
\begin{align}
&\text{QMSE}(\mathbf{a}) = \mathbb{E}\Biggl\{\Biggl\Vert\frac{\sum_{k=1}^{K}a_k\widehat{\Delta\mathbf{w}_k}+\sum_{k=1}^{K}a_k \boldsymbol\epsilon_{k}}{\eta \mathbf{1}^\top \mathbf{a}}-\frac{1}{\mathbf{1}^\top \mathbf{a}}\times\nonumber\\&\sum_{k=1}^{K}a_k\sigma_k\widehat{\Delta \mathbf{w}_k}\Biggr\Vert^2\Biggr\}= \frac{1}{(\mathbf{1}^\top\mathbf{a})^2}\Biggl(\mathbb{E}\Biggl\{\Biggl\Vert \sum_{k=1}^{K} \left(\frac{a_k}{\eta} - a_k \sigma_k\right)\times\nonumber\\
&\widehat{\Delta\mathbf{w}_k} \Biggr\Vert^2\Biggr\}+\frac{1}{\eta^2}\mathbb{E}\left\{\left\Vert\sum_{k=1}^{K}a_k \boldsymbol\epsilon_{k}\right\Vert^2\right\}\Biggr) = \frac{s}{(\mathbf{1}^\top\mathbf{a})^2}\times\nonumber\\
&\left(\left\Vert\left(\frac{1}{\eta}\mathbf{I}-\text{diag}(\boldsymbol \sigma)\right)\mathbf{a}\right\Vert^2+\frac{1}{\eta^2}\|\mathbf{a}\|^2 \sigma_\text{q}^2\right) = \frac{s}{(\mathbf{1}^\top\mathbf{a})^2}\times\nonumber\\
&\left(\mathbf{a}^\top\left(\frac{1}{\eta}\mathbf{I}-\text{diag}(\boldsymbol \sigma)\right)^2\mathbf{a}+\frac{1}{\eta^2}\|\mathbf{a}\|^2 \sigma_\text{q}^2\right),
\end{align}
where $\boldsymbol\sigma = [\sigma_1,\ldots,\sigma_K]^\top$. The factor $\eta$ that minimizes the quantization MSE is presented in the next theorem.  
\begin{theorem}
	The optimal $\eta$ for a given coefficient vector $\mathbf{a}$ is
	\begin{align}
	\eta_\text{opt} = \frac{\left(1+\sigma_\text{q}^2\right)\|\mathbf{a}\|^2}{\mathbf{a}^\top \text{diag}(\boldsymbol \sigma)\mathbf{a}}.
	\end{align}
\end{theorem}
\begin{IEEEproof}
We can expand $\frac{(\mathbf{1}^\top\mathbf{a})^2}{s} \text{QMSE}(\mathbf{a})$ as
\begin{align}
\label{qmse}
&\mathbf{a}^\top\left(\frac{1}{\eta}\mathbf{I}-\text{diag}(\boldsymbol \sigma)\right)^2\mathbf{a}+\frac{1}{\eta^2}\|\mathbf{a}\|^2 \sigma_\text{q}^2 =\nonumber\\& \mathbf{a}^\top\left(\frac{1}{\eta^2}\mathbf{I}-\frac{2}{\eta}\text{diag}(\boldsymbol{\sigma})+\text{diag}(\boldsymbol \sigma^2)\right)\mathbf{a}+\frac{1}{\eta^2}\|\mathbf{a}\|^2 \sigma_\text{q}^2.
\end{align}	
By taking derivative with respect to $\eta$ and equating the resulting expression to zero, we obtain
\begin{align}
&\mathbf{a}^\top\left(\frac{-2}{\eta^3}\mathbf{I}+\frac{2}{\eta^2}\text{diag}(\boldsymbol{\sigma})\right)\mathbf{a}-\frac{2}{\eta^3}\|\mathbf{a}\|^2 \sigma_\text{q}^2 =\nonumber\\& \frac{1}{\eta} (1+\sigma_\text{q}^2)\|\mathbf{a}\|^2 - \mathbf{a}^\top \text{diag}(\boldsymbol \sigma) \mathbf{a}= 0,
\end{align}	
which leads to the final result.
\end{IEEEproof}

Substituting the optimal $\eta$, the QMSE is
\begin{align}
\label{errorQ}
\text{QMSE}(\mathbf{a}) &= \frac{s}{(\mathbf{1}^\top\mathbf{a})^2}\left(-\frac{1}{\eta_\text{opt}}\mathbf{a}^\top \text{diag}(\boldsymbol \sigma) \mathbf{a} +\mathbf{a}^\top \text{diag}(\boldsymbol \sigma^2) \mathbf{a}\right)\nonumber\\&= \frac{s}{(\mathbf{1}^\top\mathbf{a})^2}\left(\mathbf{a}^\top \text{diag}(\boldsymbol \sigma^2) \mathbf{a} - \frac{\left(\mathbf{a}^\top \text{diag}(\boldsymbol \sigma) \mathbf{a}\right)^2}{(1+\sigma_\text{q}^2)\|\mathbf{a}\|^2}\right).
\end{align}
Fig. 3
provides a schematic representation of the receiver architecture
designed for this aggregation process. Fig. 4 illustrates a toy example of the proposed lattice coding and decoding method included in the transmission and aggregation.

\section{Convergence Analysis}
The following theorem presents the convergence analysis of {\fontfamily{lmtt}\selectfont FedCPU} in terms of the optimality gap. For perspective on the innovation of our analysis, it is worth highlighting that existing convergence studies in the field have primarily been confined to scenarios with equal aggregation weights, perfect communication conditions, or strongly convex loss functions, operating under an array of assumptions on learning parameters. In contrast, our analysis is versatile to cater to an expansive set of loss functions. It not only allows for straightforward identification of the impact of parameters and any imperfections in {\fontfamily{lmtt}\selectfont FedCPU} but also facilitates suggesting universal aggregation metrics for choosing integer coefficients, as detailed in Section V. The analysis assumes minimal common assumptions from the literature.

\textbf{Assumption 1 (Lipschitz-Continuous Gradient):} The gradient of the loss function \( F(\mathbf{w}) \) as given in \eqref{lossfunction} adheres to Lipschitz continuity characterized by a positive constant \( L \). Consequently, for any model vectors \( \mathbf{w}_1 \) and \( \mathbf{w}_2 \), the conditions below are satisfied.
\begin{align}
&F(\mathbf{w}_2) \leq F(\mathbf{w}_1) + \nabla F(\mathbf{w}_1)^T (\mathbf{w}_2-\mathbf{w}_1) + \frac{L}{2} \|\mathbf{w}_2 - \mathbf{w}_1\|^2,\\
&\|\nabla F(\mathbf{w}_2)-\nabla F(\mathbf{w}_1)\| \leq L \|\mathbf{w}_2 - \mathbf{w}_1\|.
\end{align}


\textbf{Assumption 2 (Gradient Variance Bound):} At a device, the local gradient estimate \( \mathbf{g} \) serves as an unbiased estimate of the ground-truth gradient \( \nabla F(\mathbf{w}) \), and its variance remains bounded, as
\begin{align}
\mathbb{E}\left\{\|\mathbf{g} - \nabla F(\mathbf{w})\|^2\right\} \leq \frac{\sigma_\text{g}^2}{B}.
\end{align}

\textbf{Assumption 3 (Polyak-Lojasiewicz Inequality):} Let \( F^* = F(\mathbf{w}^*) \) be from problem \eqref{objective}. There exists a constant \( \delta \geq 0 \) for which the subsequent condition holds.
\begin{align}
\label{polyak}
\|\nabla F(\mathbf{w})\|^2 \geq 2\delta \left(F(\mathbf{w}) - F^*\right).
\end{align}
The inequality presented in \eqref{polyak} is significantly more expansive and general than the mere assumption of convexity \cite{karimi}.
\begin{theorem}
Let $1-\frac{L^2\mu_t^2}{2}\tau(\tau-1)-L\mu_t \tau \geq 0$, $1-\mu_t \tau \delta \geq 0$, and $\mathbf{a}_t$ as the integer coefficient vector for each round $ t\in\left\{0,\ldots,T-1\right\}$, in the absence of decoding error, i.e., $P_{\text{D},t} \to 0$, we have
\begin{align}
&\mathbb{E}\left\{F({\mathbf{w}}_{\text{G},T})\right\}-F^* \leq \left(\prod_{t=0}^{T-1} c_t\right) \biggl(\mathbb{E}\left\{F({\mathbf{w}}_{\text{G},0})\right\}-F^*\biggr)+\nonumber\\&\sum_{t=0}^{T-1}\left(b_t+\frac{L}{2}{\cal L}_t(\mathbf{a}_t)\right)\prod_{i=t+1}^{T-1}c_i,
\end{align}
where
\begin{align}
c_t = 1-\mu_t \tau \delta,
\end{align}
\begin{align}
b_t = \frac{L^2\mu_{t}^3}{2}\frac{\tau (\tau-1)}{2}\frac{\sigma_\text{g}^2}{B},
\end{align}
\begin{align}
\label{mainL}
{\cal L}_t(\mathbf{a}_t) = { \mu_t^2}\frac{\sigma_\text{g}^2}{B}\tau \frac{\|\mathbf{a}_t\|^2}{(\mathbf{1}^\top\mathbf{a}_{t})^2} + \text{QMSE}_t(\mathbf{a}_t),
\end{align}
where $\text{QMSE}_t$ is given in \eqref{errorQ}.
\end{theorem}
\begin{IEEEproof}
	See Appendix A.
\end{IEEEproof}
\begin{remark}
The optimality gap encompasses both learning and communication elements like quantization error, decoding error, integer coefficients of the decoded lattice point, learning rate, and batch size. This underscores the holistic approach of {\fontfamily{lmtt}\selectfont FedCPU} which melds communication and learning in its design.
\end{remark}

From Theorem 3 and as a special case of {\fontfamily{lmtt}\selectfont FedCPU}, the optimality gap for {\fontfamily{lmtt}\selectfont FedAvg} when operating under error-free
transmission and ideal aggregation \eqref{agg} is given in the following. 
\begin{corollary}
Let $1-\frac{L^2\mu_t^2}{2}\tau(\tau-1)-L\mu_t \tau \geq 0$, $1-\mu_t \tau \delta \geq 0$, and $\mathbf{a}_t = \mathbf{1}$ for each round $ t\in\left\{0,\ldots,T-1\right\}$, then in the case of error-free transmission, we have
\begin{align}
&\mathbb{E}\left\{F({\mathbf{w}}_{\text{G},T})\right\}-F^* \leq \left(\prod_{t=0}^{T-1} c_t\right) \biggl(\mathbb{E}\left\{F({\mathbf{w}}_{\text{G},0})\right\}-F^*\biggr)+\nonumber\\&\sum_{t=0}^{T-1}\left(b_t+\frac{L}{2}{\cal L}_t\right)\prod_{i=t+1}^{T-1}c_i,
\end{align}
where
\begin{align}
\label{mainI}
{\cal L}_t = { \mu_t^2}\frac{\sigma_\text{g}^2}{B}\tau \frac{1}{K}.
\end{align}
\end{corollary}
\begin{remark}
Errors in {\fontfamily{lmtt}\selectfont FedCPU} contribute to an increase in the ${\cal L}_t$ term by ${ \mu_t^2}\frac{\sigma_\text{g}^2}{B}\tau \Bigl(\frac{\|\mathbf{a}_t\|^2}{(\mathbf{1}^\top\mathbf{a}_{t})^2}-\frac{1}{K}\Bigr) + \text{QMSE}_t(\mathbf{a}_t)$. This augmentation arises not only from the MSE due to the quantization error but also from the mismatch between the coefficient vector $\mathbf{a}_t$ and $\mathbf{1}$, where the latter corresponds to the ideal aggregation in \eqref{agg}. The mismatch arises from imperfections on the learning aspect of {\fontfamily{lmtt}\selectfont FedCPU}.
\end{remark}
\begin{remark}
The mismatch term $\frac{\|\mathbf{a}_t\|^2}{(\mathbf{1}^\top\mathbf{a}_{t})^2}-\frac{1}{K}$ is amplified by the factor ${ \mu_t^2}\frac{\sigma_\text{g}^2}{B}\tau$. This implies that compared to the MSE the impact of the mismatch becomes more significant with larger values of the local training parameters $\mu_t$ and $\tau$. Notably, $\mu_t$ exerts a greater impact.
\end{remark}

\begin{remark}
When decoding error is taken into account in the analysis, another term is added to 
${\cal L}_t(\mathbf{a}_t)$ in \eqref{mainL}. Characterizing this term is mathematically intractable. For each lattice point, which has a Voronoi region with a complex shape, it is necessary to first compute the conditional probability that the decoding noise, with variance 
$\text{DMSE}$ as in \eqref{dmsee}, lies within this region. Subsequently, we need to compute the distance between this lattice point and the target one.
\end{remark}

\begin{remark}
In the absence of error, the minimum of the term \(\mathcal{L}_t(\mathbf{a}_t)\) is achieved when \(\mathbf{a}_t = \mathbf{1}\), as this results in the mismatch term becoming zero. However, in the practical scenario of {\fontfamily{lmtt}\selectfont FedCPU} involving quantization and imperfect wireless communications, characterized by interference and noise, the optimal aggregation that minimizes \(\mathcal{L}_t(\mathbf{a}_t)\) while still accommodating a small decoding error diverges from the ideal aggregation.
\end{remark}

\section{Integer Coefficient Selection}
For the selection of integer coefficient vectors \( \mathbf{a}_t \) for all \( t \in \left\{0,\ldots,T-1\right\} \), inspired by \textit{Remark 7}, we aim to minimize the optimality gap as described in Theorem 3. Equally, this minimization can be broken down into individual subproblems, each corresponding to one round \( t \). 
\begin{align}
\label{main_obj}
&\mathbf{a}_t = \arg\min_{\mathbf{a}\in \mathbb{Z}^{K \times 1}\backslash \left\{\mathbf{0}\right\}} {\cal L}_t(\mathbf{a}) = { \mu_t^2}\frac{\sigma_\text{g}^2}{B}\tau \frac{\|\mathbf{a}\|^2}{(\mathbf{1}^\top\mathbf{a})^2} + \text{QMSE}_t(\mathbf{a})\nonumber\\&={ \mu_t^2}\frac{\sigma_\text{g}^2}{B}\tau \frac{\|\mathbf{a}\|^2}{(\mathbf{1}^\top\mathbf{a})^2} \nonumber\\&+ \frac{s}{(\mathbf{1}^\top\mathbf{a})^2}\left(\mathbf{a}^\top \text{diag}(\boldsymbol \sigma_t^2) \mathbf{a} - \frac{\left(\mathbf{a}^\top \text{diag}(\boldsymbol \sigma_t) \mathbf{a}\right)^2}{(1+\sigma_\text{q}^2)\|\mathbf{a}\|^2}\right). 
\end{align}
subject to
\begin{equation}
\begin{cases}
a_k \geq 0, \ \forall k,\\
P_{\text{D},t} \leq \varepsilon,
\end{cases}\nonumber
\end{equation}
where $\varepsilon > 0$ is small enough. In \eqref{main_obj}, the function ${\cal L}_t(\mathbf{a})$, which combines both the learning and communication characteristics of {\fontfamily{lmtt}\selectfont FedCPU}, acts as the selection metric for $\mathbf{a}$. Given that $\mathbf{a}$ sets the aggregation weights in \eqref{model_estimate}, this metric can be termed the aggregation metric. The optimization problem in \eqref{main_obj} is an integer programming that falls under the NP-hard category. Moreover, to determine the gradient variance bound $\sigma_\text{g}^2$, one must understand the gradient data statistics--- which is unavailable in many applications. Hence, considering our aim for universality, we propose an alternative problem which does not depend on this specific knowledge.\footnote{Unlike our approach and many in FL which do not assume this knowledge, some studies, like \cite{cao}, do or estimate it. In such cases, future research has the potential to address the primary problem \eqref{main_obj}.} 

Given that the standard deviation of model updates across devices does not vary significantly in most learning scenarios, i.e., $\sigma_{i,t} \approx \sigma_{j,t}, j \neq i$, also noted in \cite{gunduz2, cao}, we can approximate $\text{QMSE}_t(\mathbf{a})$ as
\begin{align}
&\text{QMSE}_t(\mathbf{a}) = \frac{s}{(\mathbf{1}^\top\mathbf{a})^2}\left(\mathbf{a}^\top \text{diag}(\boldsymbol \sigma_t^2) \mathbf{a} - \frac{\left(\mathbf{a}^\top \text{diag}(\boldsymbol \sigma_t) \mathbf{a}\right)^2}{(1+\sigma_\text{q}^2)\|\mathbf{a}\|^2}\right) \nonumber
\end{align}
\begin{align}
&\approx \frac{s}{(\mathbf{1}^\top\mathbf{a})^2} \left(\sigma_{1,t}^2 -\frac{\sigma_{1,t}^2}{1+\sigma_\text{q}^2}\right)\|\mathbf{a}\|^2 = s \sigma_{1,t}^2 \frac{\sigma_\text{q}^2}{1+\sigma_\text{q}^2} \frac{\|\mathbf{a}\|^2}{(\mathbf{1}^\top\mathbf{a})^2},
\end{align}
which leads to
\begin{align}
{\cal L}_t(\mathbf{a}) &\approx \left({ \mu_t^2}\frac{\sigma_\text{g}^2}{B}\tau + s \sigma_{1,t}^2 \frac{\sigma_\text{q}^2}{1+\sigma_\text{q}^2}\right) \frac{\|\mathbf{a}\|^2}{(\mathbf{1}^\top\mathbf{a})^2}.
\end{align}
Thus, it is enough to solve the following optimization problem
\begin{align}
\label{integer_opt}
\mathbf{a}_t=\arg\min_{\mathbf{a}\in \mathbb{Z}^{K \times 1}\backslash \left\{\mathbf{0}\right\}} \frac{\|\mathbf{a}\|^2}{(\mathbf{1}^\top\mathbf{a})^2},
\end{align}
subject to
\begin{equation}
\begin{cases}
({1+2\sigma_\text{q}^2})\mathbf{a}^\top\left( \mathbf{I} +{\text{SNR}}\mathbf{H}_t^\top\mathbf{H}_t\right)^{-1} \mathbf{a}\leq \theta,\\
a_k \geq 0, \ \forall k,
\end{cases}\nonumber
\end{equation}
where the threshold $\theta$ is chosen to ensure that the decoding MSE, the DMSE in \eqref{dmsee}, remains sufficiently small, i.e., the decoding noise lies within $\mathbf{a}^\top \overline{\Delta\mathbf{W}}+\mathcal{V}$ with high probability or $P_{\text{D},t}$ is small enough. In \eqref{integer_opt}, the objective minimization has a dual target as it simultaneously minimizes both the mismatch and the quantization MSE. 

The integer programming problem \eqref{integer_opt} is still complicated and solving it poses a considerable challenge. This could be an interesting area for future research. In the meantime, we suggest a suboptimal solution by introducing the following problem.
\begin{align}
\label{real_opt}
\mathbf{a}_t= \text{round-to-integer}\left\{\arg\min_{\mathbf{a}\in \mathbb{R}^{K \times 1}} \frac{\|\mathbf{a}\|^2}{(\mathbf{1}^\top\mathbf{a})^2}\right\},
\end{align}
subject to
\begin{equation}
\begin{cases}
\mathbf{a}^\top\left( \mathbf{I} +{\text{SNR}}\mathbf{H}_t^\top\mathbf{H}_t\right)^{-1}\mathbf{a} \leq \frac{\theta}{{1+2\sigma_\text{q}^2}},\\
a_k \geq 1, \ \forall k.
\end{cases}\nonumber
\end{equation}
In the absence of error, \eqref{real_opt} excludes its first constraint, making it evident that the solution is $\mathbf{a}_t = \mathbf{1}$. Since \eqref{real_opt} is non-convex, we recommend a successive convexification strategy. In this approach, for each iteration denoted by $n$, we address the corresponding convex problem as
\begin{align}
\label{real_convex}
\mathbf{a}_t^{(n)}= \arg\min_{\mathbf{a}\in \mathbb{R}^{K \times 1}} {\|\mathbf{a}\|^2},
\end{align}
subject to
\begin{equation}
\begin{cases}
\mathbf{1}^\top \mathbf{a} = \mathbf{1}^\top \mathbf{a}_t^{(n-1)},\\
\mathbf{a}^\top\left( \mathbf{I} +{\text{SNR}}\mathbf{H}_t^\top{\mathbf{H}_t}\right)^{-1}\mathbf{a} \leq \frac{\theta}{{1+2\sigma_\text{q}^2}},\\
a_k \geq 1, \ \forall k.
\end{cases}\nonumber
\end{equation}
This implies that after $N$ iterations, the solution is $\mathbf{a}_t = \text{round-to-integer}\left\{\mathbf{a}_t^{(N)}\right\}$. The problem \eqref{real_convex} can be efficiently solved using widely employed convex optimization tools. Despite the approximations between \eqref{main_obj} and \eqref{real_convex}, we demonstrate in Section VI that the experimental results can still closely achieve the ideal performance with no transmission errors.  

Before ending this section, it is important to mention that if devices have access to CSIT, they can optimize their transmission powers, which can further decrease the decoding MSE for a given coefficient vector $\mathbf{a}$, i.e., 
$\text{DMSE}(\mathbf{a})$, thereby improving the learning performance.

\section{Experimental Results}
The task involves classifying images from two datasets of standard MNIST and Fashion-MNIST with the parameter
values given in Table 1, unless otherwise stated. We have constructed the classifier using a Convolutional Neural Network (CNN). This CNN features two convolutional layers, both of $3 \times 3$ size with ReLU activation--- the first containing 32 channels and the second 64. Each of these layers is succeeded by a $2 \times 2$ max pooling. Subsequent layers include a fully connected layer with 128 units with ReLU activation, culminating in a softmax output layer. The lattice generator matrix is given by \(\mathbf{G} = \text{diag}\left\{\mathbf{G}_8, \dots, \mathbf{G}_8\right\}\), indicating an orthogonal construction with size \(s\), where \(\mathbf{G}_8\) represents the generator matrix of the E8 lattice \cite{e8}. Both i.i.d. and non-i.i.d. distributions of dataset samples among devices are considered. For the non-i.i.d. scenario, each device contains samples exclusively from two classes, and the sample count differs from one device to another. Performance is gauged by the learning accuracy in relation to the test dataset throughout the global iteration count, denoted by $t$. The outcome for performance is determined by averaging 20 realization samples. This ensures Gaussian channel distribution is considered--- specifically, channel gain with Rayleigh fading $\sim \exp(5)$ and the channel phase following the uniform distribution $\sim {\cal{U}}(0,2{\pi})$.

\begin{table}
	\caption {Parameter Values} 
	\vspace{-8pt}
	\begin{center}
		\resizebox{4.5cm}{!} {
			\begin{tabular}{| l | l | l | l | l | l | l | l | l | l | l | l | l | l}
				
				\hline
				\hline
				{$K$}&{$\text{SNR}$}&{$\tau$}& $\mu$& $B$&$M$ \\ \hline
				$30$& $10$ &$3$&$0.01$& $100$&$30$  \\ \hline	
				\hline
		\end{tabular}}
	\end{center}
	\vspace{-7pt}
\end{table}

\begin{figure}[tb!]
	\vspace{0pt}
	\centering
	\includegraphics[width =2.6in]{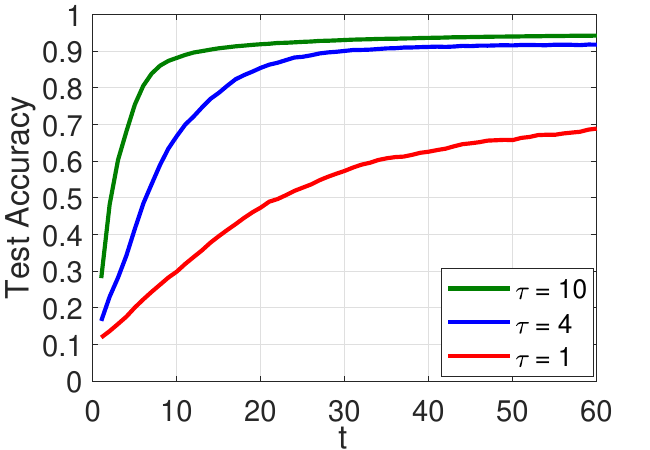} 
	\vspace{-5pt}
	\caption{MNIST, i.i.d. case.}
	\vspace{-5pt}
\end{figure}

\begin{figure}[tb!]
	\vspace{0pt}
	\centering
	\includegraphics[width =2.6in]{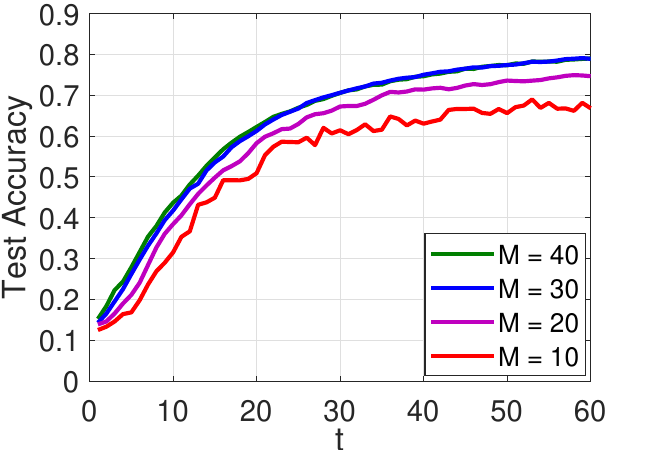} 
	\vspace{-5pt}
	\caption{MNIST, non-i.i.d. case.}
	\vspace{-5pt}
\end{figure}

\begin{figure}[tb!]
	\vspace{0pt}
	\centering
	\includegraphics[width =2.6in]{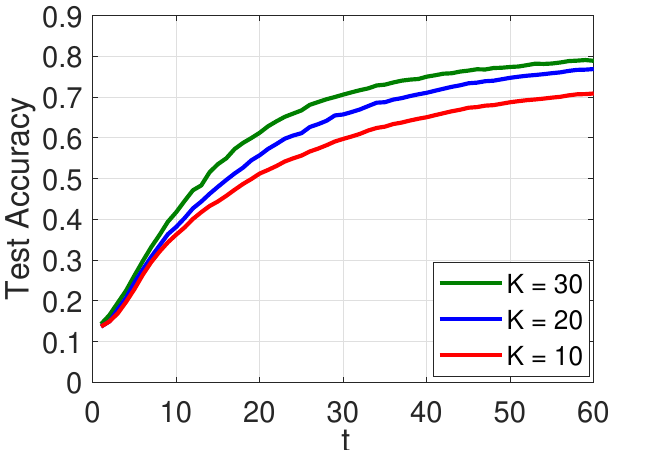} 
	\vspace{-5pt}
	\caption{MNIST, non-i.i.d. case.}
	\vspace{-5pt}
\end{figure}

\begin{figure}[tb!]
	\vspace{0pt}
	\centering
	\includegraphics[width =2.6in]{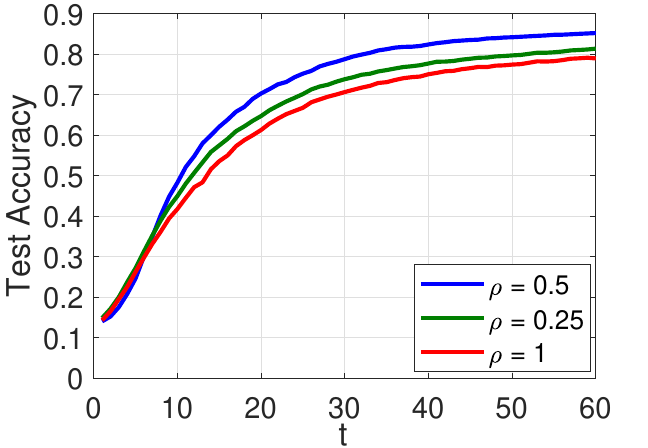} 
	\vspace{-5pt}
	\caption{MNIST, non-i.i.d. case.}
	\vspace{-5pt}
\end{figure}

\begin{figure}[tb!]
	\vspace{0pt}
	\centering
	\includegraphics[width =2.6in]{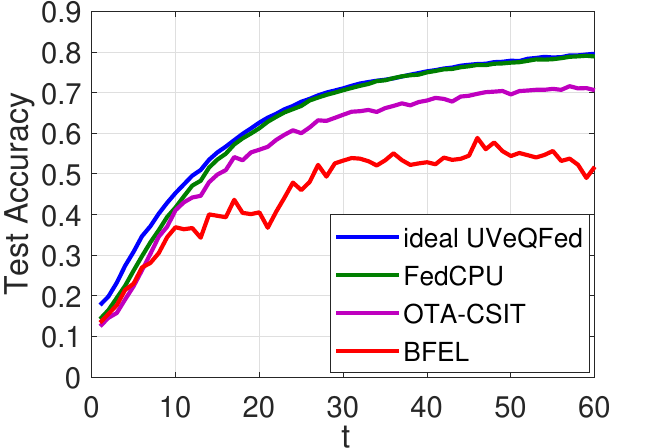} 
	\vspace{-5pt}
	\caption{MNIST, non-i.i.d. case.}
	\vspace{-5pt}
\end{figure}

\begin{figure}[tb!]
	\vspace{0pt}
	\centering
	\includegraphics[width =2.6in]{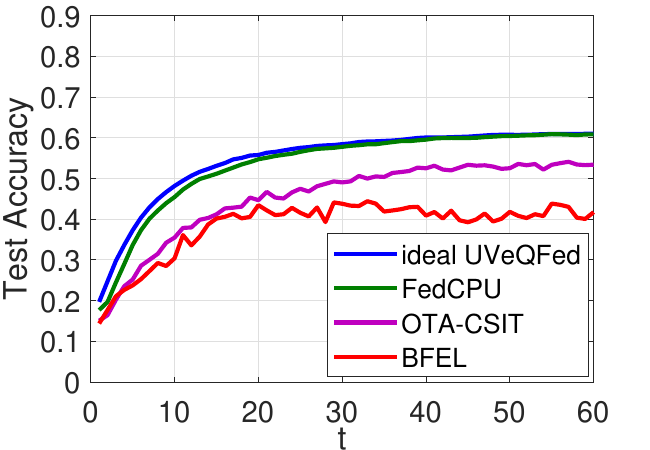} 
	\vspace{-5pt}
	\caption{Fashion-MNIST, non-i.i.d. case.}
	\vspace{-5pt}
\end{figure}

\begin{figure}[tb!]
	\vspace{0pt}
	\centering
	\includegraphics[width =2.6in]{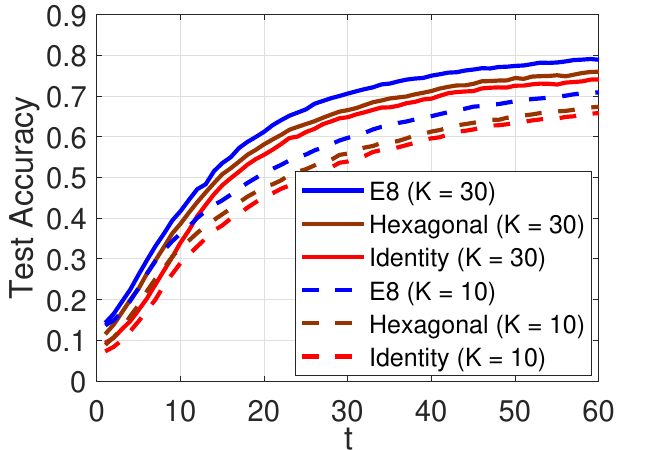} 
	\vspace{-5pt}
	\caption{MNIST, non-i.i.d. case.}
	\vspace{-5pt}
\end{figure}

In Fig. 5, the accuracy is shown for different local iterations $\tau$ in the MNIST and i.i.d.\ scenario. It is evident that by augmenting $\tau$ or $t$, there is a positive impact on learning performance. Additionally, the performance sees a significant boost when integrating multiple local iterations instead of relying on a single one.

In Fig. 6, the accuracy is shown for different numbers of antennas $M$ at the server in the MNIST and non-i.i.d. scenario. The performance improves as $M$ increases because the decoding MSE is decreased. However, for higher values of $M$, this performance improvement tends to diminish.

Fig. 7 displays the accuracy for different numbers of devices $K$ in the MNIST and non-i.i.d. scenario. As $K$ increases, accuracy improves due to more devices participating in the learning process. However, this improvement comes at the cost of increased decoding MSE, creating a tradeoff. Specifically, for higher values of $K$, the performance gains in accuracy tend to diminish.

In Fig. 8, we explore the implications of lattice quantization by employing different lattice generator matrices as 
$\rho \mathbf{G}$ for the MNIST and non-i.i.d. scenario. As 
$\rho$ decreases, the lattice points become more densely packed, leading to smaller Voronoi regions. Notably, while a reduced 
$\rho$ minimizes the quantization error, it does not necessarily enhance performance due to the rise in decoding error. This results in a balance of factors; strikingly, performance sees a marked improvement at $\rho = 0.5$ in comparison to $\rho = 0.25$ and $\rho = 1$. 

In Figs. 9 and 10, comparisons are made between {\fontfamily{lmtt}\selectfont
	FedCPU} and schemes from the literature, serving as benchmarks, for non-i.i.d. scenario and both MNIST and Fashion-MNIST. Among the evaluated benchmarks are {\fontfamily{lmtt}\selectfont ideal UVeQFed} \cite{eldar}, {\fontfamily{lmtt}\selectfont OTA-CSIT} \cite{ding}, and {\fontfamily{lmtt}\selectfont BFEL} \cite{gunduz4} schemes. The {\fontfamily{lmtt} \selectfont ideal UVeQFed} is the version of {\fontfamily{lmtt} \selectfont UVeQFed} when operating with ideal communications, given that {\fontfamily{lmtt} \selectfont UVeQFed} does not delve into the communication aspect of FL. It employs lattice quantization and leverages orthogonal transmission to completely avoid interference, operating under the premise of boundless communication resources. Also, {\fontfamily{lmtt}\selectfont ideal UVeQFed} benefits from robust channel coding techniques, ensuring error-free decoding to counteract channel noise. Please note that the superiority of {\fontfamily{lmtt}\selectfont UVeQFed} compared to other quantization methods is justified in \cite{eldar}. The same lattice generator matrix $\mathbf{G}$ is considered for the {\fontfamily{lmtt}\selectfont ideal UVeQFed}. While comparing the accuracy of our over-the-air scheme with an orthogonal scheme like {\fontfamily{lmtt} \selectfont UVeQFed}--- which focuses solely on quantization noise and demands a minimum of $K$ times additional resources, disregarding the resources for extended channel coded sequences--- might seem unfair, we initiate this comparison to showcase the capability of {\fontfamily{lmtt}\selectfont
	FedCPU} in reaching the benchmarks set by such an ideal model. In contrast, the {\fontfamily{lmtt}\selectfont OTA-CSIT} and {\fontfamily{lmtt}\selectfont BFEL} schemes stand out as over-the-air benchmarks that rely on multi-antenna processing at the server to mitigate interference effects, doing so without employing quantization. {\fontfamily{lmtt}\selectfont BFEL}, characterized as a blind approach, transmits at a constant power, whereas {\fontfamily{lmtt}\selectfont OTA-CSIT}, benefiting from CSIT, implements power control at the devices to enhance system performance. Different from {\fontfamily{lmtt}\selectfont FedCPU} and {\fontfamily{lmtt}\selectfont BFEL}, which rely exclusively on CSIR, {\fontfamily{lmtt}\selectfont OTA-CSIT} necessitates both CSIR and CSIT as well as additional hardware on each device for channel compensation. For all three schemes, {\fontfamily{lmtt}\selectfont FedCPU}, {\fontfamily{lmtt}\selectfont OTA-CSIT}, and {\fontfamily{lmtt}\selectfont BFEL}, the number of antennas at the server is set to $M= 30$.

The performance superiority of {\fontfamily{lmtt}\selectfont FedCPU} over {\fontfamily{lmtt}\selectfont OTA-CSIT} and {\fontfamily{lmtt}\selectfont BFEL} is evident. This improvement is attributed to the implementation of adaptive aggregation weights, a component that was previously not given due consideration. Furthermore, {\fontfamily{lmtt}\selectfont FedCPU} incorporates an innovative equalization method paired with a unique receiver design at the server, a strategy that differs from the one outlined in \cite{gunduz4}. Also, {\fontfamily{lmtt}\selectfont OTA-CSIT}, constrained by average and maximum power limits, requires selective device participation. On the other hand, {\fontfamily{lmtt}\selectfont FedCPU} allows all devices to contribute, each with an effective aggregation weight. Additionally, {\fontfamily{lmtt}\selectfont FedCPU}, subjected to interference and noise, nearly matches the performance of {\fontfamily{lmtt}\selectfont ideal UVeQFed}, which operates without interference and noise. This is because, with the same quantization error assumed for both schemes using the same lattice code, {\fontfamily{lmtt}\selectfont FedCPU}--- leveraging multi-antenna processing, weighted aggregation, and optimal receiver design at the server--- brings decoding errors close to zero, similar to the {\fontfamily{lmtt}\selectfont ideal UVeQFed}.

In Fig. 11, the impact of various lattice codes within {\fontfamily{lmtt}\selectfont
		FedCPU} on accuracy is evaluated with $K=10$ and 30 for the MNIST and non-i.i.d.\ scenario. The orthogonal construction is applied using a generator matrix of the form \(\mathbf{G} = \text{diag}\left\{\mathbf{G}_n, \dots, \mathbf{G}_n\right\}\). The codes considered include the identity lattice with \(\text{G}_1 = 1\), the hexagonal lattice with \(\mathbf{G}_2 = \begin{bmatrix} 1 & 0 \\ \frac{1}{2} & \frac{\sqrt{3}}{2} \end{bmatrix}\), and the E8 lattice with $\mathbf{G}_8$ as in \cite{e8}. The results indicate that these well-designed lattice codes perform better as their dimensionality increases, due to a reduction in the lattice's second moment, which in turn decreases quantization and decoding errors. However, it should be noted that the complexity of implementation and decoding increases accordingly.

\section{Conclusions}
We introduced a federated
learning scheme incorporating lattice codes, pioneering a new over-the-air computation method. The proposed scheme offers adjustable quantization, enabling distributed learning through digital modulation. Additionally, it ensures resilience against interference and noise through coding. In this scheme, with no need for channel compensation at the transmitter end, an integer combination of lattice quantized model parameters is reliably decoded and processed for aggregation.
We derived the optimality gap for the learning process within the scheme, contingent on the integer coefficients and encompassing both communication and learning factors. To determine these integer coefficients, we have suggested a tailored aggregation metric rooted in the gap. For the optimization of the metric, which falls under NP-hard integer programming, we proposed a method for convexification and presented an efficient
algorithm. In spite of channel conditions and data heterogeneity, experimental findings showcased the superior learning accuracy of the proposed scheme, outperforming existing over-the-air alternatives. Furthermore, even with a limited number of antennas at the server, the proposed scheme can nearly achieve the performance level anticipated in a scenario devoid of transmission errors.

\appendices
\section{Proof of Theorem 1}
The update for the learning model at round $t+1$ is represented as
\begin{align} &\Delta{\mathbf{w}_{\text{G},t+1}} = {\mathbf{w}}_{\text{G},t+1} - \mathbf{w}_{\text{G},t}= \frac{1}{\mathbf{1}^\top \mathbf{a}_t} \sum_{k=1}^{K}a_{k,t} \Delta \mathbf{w}_{k,t}^\top+{\boldsymbol\epsilon_\text{Q}(\mathbf{a}_{t})}\nonumber\\
&=  -\frac{\mu_{t}}{\mathbf{1}^\top\mathbf{a}_{t}}\sum_{k=1}^{K}a_{k,t} \sum_{i=0}^{\tau-1}\nabla F_k(\mathbf{w}_{k,t,i},\boldsymbol\xi_{k}^i)+{\boldsymbol\epsilon_\text{Q}(\mathbf{a}_{t})},
\end{align}
where $\boldsymbol\epsilon_\text{Q}(\mathbf{a}_t)$ is the quantization error in recovering $\mathbf{a}_t$ as
\begin{align}
\label{errorqqq}
&\boldsymbol\epsilon_\text{Q}(\mathbf{a}_t) = \frac{\sum_{k=1}^{K}a_{k,t}\widehat{\Delta\mathbf{w}_{k,t}}^\top+\sum_{k=1}^{K}a_{k,t} \boldsymbol\epsilon_{k,t}^\top}{\eta_t \mathbf{1}^\top \mathbf{a}_t}-\nonumber\\
&\frac{1}{\mathbf{1}^\top \mathbf{a}_t}\sum_{k=1}^{K}a_{k,t}\sigma_{k,t}\widehat{\Delta \mathbf{w}_{k,t}}^\top.
\end{align}
Then, based on the $L$-Lipschitz continuous property outlined in Assumption 1, we can infer
\begin{align}
\label{lipsch}
&F({\mathbf{w}}_{\text{G},t+1}) - F({\mathbf{w}}_{\text{G},t}) \leq \nabla F( {\mathbf{w}}_{\text{G},t})^\top \left({\mathbf{w}}_{\text{G},t+1} -  {\mathbf{w}}_{\text{G},t}\right)+\frac{L}{2}\times\nonumber\\
& \| {\mathbf{w}}_{\text{G},t+1} - {\mathbf{w}}_{\text{G},t}\|^2 =\nabla F( {\mathbf{w}}_{\text{G},t})^\top \Biggl(-\frac{\mu_{t}}{\mathbf{1}^\top\mathbf{a}_{t}}\sum_{k=1}^{K}a_{k,t}\nonumber\\
& \sum_{i=0}^{\tau-1}\nabla F_k(\mathbf{w}_{k,t,i},\boldsymbol\xi_{k}^i)+\boldsymbol\epsilon_\text{Q}(\mathbf{a}_t)\Biggr)+\frac{L}{2} \Biggl\Vert -\frac{\mu_{t}}{\mathbf{1}^\top\mathbf{a}_{t}}\sum_{k=1}^{K}a_{k,t}\nonumber\\
& \sum_{i=0}^{\tau-1}\nabla F_k(\mathbf{w}_{k,t,i},\boldsymbol\xi_{k}^i)+\boldsymbol\epsilon_\text{Q}(\mathbf{a}_t)\Biggr\Vert^2.
\end{align}
By taking the expectation on both sides of \eqref{lipsch}, we continue as
\begin{align}
\label{expected}
&\mathbb{E}\left\{F( {\mathbf{w}}_{\text{G},t+1}) - F( {\mathbf{w}}_{\text{G},t}) \right\} \leq -\frac{\mu_{t}}{\mathbf{1}^\top\mathbf{a}_{t}}\sum_{k=1}^{K}a_{k,t} \nonumber\\
&\sum_{i=0}^{\tau-1} \mathbb{E}\left\{\nabla F( {\mathbf{w}}_{\text{G},t})^\top \nabla F_k(\mathbf{w}_{k,t,i},\boldsymbol\xi_{k}^i)\right\}+\frac{L\mu_t^2}{2} \mathbb{E}\Biggl\{ \Biggl\Vert \sum_{k=1}^{K}\frac{a_{k,t}}{\mathbf{1}^\top\mathbf{a}_{t}} \nonumber\\
&\sum_{i=0}^{\tau-1}\nabla F_k(\mathbf{w}_{k,t,i},\boldsymbol\xi_{k}^i)\Biggr\Vert^2\Biggr\} +\frac{L}{2}\mathbb{E}\left\{\left\Vert \boldsymbol\epsilon_\text{Q}(\mathbf{a}_t)\right\Vert^2\right\}= -\frac{\mu_{t}}{\mathbf{1}^\top\mathbf{a}_{t}}\nonumber\\&\sum_{k=1}^{K}a_{k,t} \sum_{i=0}^{\tau-1} \mathbb{E}\left\{\nabla F( {\mathbf{w}}_{\text{G},t})^\top \nabla F(\mathbf{w}_{k,t,i})\right\}+\frac{L\mu_t^2}{2} \times\nonumber\\
&\mathbb{E}\left\{ \left\Vert \sum_{k=1}^{K}\frac{a_{k,t}}{\mathbf{1}^\top\mathbf{a}_{t}} \sum_{i=0}^{\tau-1}\nabla F_k(\mathbf{w}_{k,t,i},\boldsymbol\xi_{k}^i)\right\Vert^2\right\} +\frac{L}{2} {\text{QMSE}_t(\mathbf{a}_{t})},
\end{align}
where $\mathbb{E}\left\{\boldsymbol\epsilon_\text{Q}(\mathbf{a}_t)\right\} = \mathbf{0}$ is replaced, which comes from $\mathbb{E}\left\{\widehat{\Delta\mathbf{w}_{k,t}}\right\} = \mathbf{0}$ and $\mathbb{E}\left\{\boldsymbol\epsilon_{k,t}\right\} = \mathbf{0}, \forall k$ in \eqref{errorqqq}. Then, we bound the first term of the right side in \eqref{expected}. By employing the equation $\|\mathbf{t}_1-\mathbf{t}_2\|^2 = \|\mathbf{t}_1\|^2+\|\mathbf{t}_2\|^2- 2\mathbf{t}_1^\top \mathbf{t}_2$ for any vectors $\mathbf{t}_1$ and $\mathbf{t}_2$, we express its inner-sum term as
\begin{align}
\label{diffexp}
&\mathbb{E}\left\{\nabla F( {\mathbf{w}}_{\text{G},t})^\top \nabla F(\mathbf{w}_{k,t,i})\right\} = \frac{1}{2} \mathbb{E}\left\{\|\nabla F( {\mathbf{w}}_{\text{G},t})\|^2\right\} + \frac{1}{2}\times \nonumber\\
&\mathbb{E}\left\{\|\nabla F(\mathbf{w}_{k,i,t})\|^2\right\} - \frac{1}{2} \mathbb{E}\left\{\|\nabla F( {\mathbf{w}}_{\text{G},t})- \nabla F(\mathbf{w}_{k,i,t})\|^2\right\}.
\end{align}
Based on Assumption 1, the last term in \eqref{diffexp} can be bounded as
\begin{align}
\label{diffnorm}
&\mathbb{E}\left\{\|\nabla F( {\mathbf{w}}_{\text{G},t})- \nabla F(\mathbf{w}_{k,i,t})\|^2\right\} \leq L^2 \mathbb{E}\left\{\|\mathbf{w}_{\text{G},t}-\mathbf{w}_{k,i,t}\|^2\right\}\nonumber\\&= L^2 \mathbb{E}\left\{\left\Vert-\mu_t \sum_{j=0}^{i-1} \nabla F_k(\mathbf{w}_{k,t,j},\boldsymbol\xi_{k}^j)\right \Vert^2\right\}\nonumber\\
&=L^2\mu_t^2 \mathbb{E}\left\{\left\Vert\sum_{j=0}^{i-1}\nabla F_k(\mathbf{w}_{k,t,j},\boldsymbol\xi_{k}^j) \right\Vert^2\right\},
\end{align}
where using the identity $\mathbb{E}\left\{\|\mathbf{t}\|^2\right\} = \|\mathbb{E}\left\{\mathbf{t}\right\}\|^2 + \mathbb{E}\left\{\|\mathbf{t}-\mathbb{E}\left\{\mathbf{t}\right\}\|^2\right\}$ for any vector $\mathbf{t}$, we obtain
\begin{align}
\label{normexp}
&\mathbb{E}\left\{\left\Vert\sum_{j=0}^{i-1}\nabla F_k(\mathbf{w}_{k,t,j},\boldsymbol\xi_{k}^j) \right\Vert^2\right\} =\mathbb{E}\left\{\left\Vert\sum_{j=0}^{i-1}\nabla F(\mathbf{w}_{k,t,j}) \right\Vert^2\right\}\nonumber\\
&+\mathbb{E}\left\{\left\Vert \sum_{j=0}^{i-1} \nabla F_k(\mathbf{w}_{k,j,t},\boldsymbol\xi_{k}^j)- \nabla F(\mathbf{w}_{k,j,t})\right\Vert^2\right\},
\end{align}
where the first term on the right side can be expressed with an upper bound as
\begin{align}
\label{RHS1}
\mathbb{E}\left\{\left\Vert\sum_{j=0}^{i-1}\nabla F(\mathbf{w}_{k,t,j}) \right\Vert^2\right\} &\stackrel{(a)}{\leq}  i \sum_{j=0}^{i-1}\mathbb{E}\left\{\left\Vert\nabla F(\mathbf{w}_{k,t,j}) \right\Vert^2\right\},
\end{align}
where $(a)$ follows from the arithmetic-geometric mean inequality, given by 
\(
\left(\sum_{j=1}^{J}t_j\right)^2 \leq J \sum_{j=1}^{J}t_j^2.
\)
The second term on the right side of \eqref{normexp} can be upper bounded as follows
\begin{align}
\label{RHS2}
&\mathbb{E}\left\{\left\Vert \sum_{j=0}^{i-1} \nabla F_k(\mathbf{w}_{k,j,t},\boldsymbol\xi_{k}^j)- \nabla F(\mathbf{w}_{k,j,t})\right\Vert^2\right\}\stackrel{(b)}{=} \nonumber\\
& \sum_{j=0}^{i-1}\mathbb{E}\left\{\left\Vert \nabla F_k(\mathbf{w}_{k,j,t},\boldsymbol\xi_{k}^j)- \nabla F(\mathbf{w}_{k,j,t})\right\Vert^2\right\}
\stackrel{(c)}{\leq} i\frac{\sigma_\text{g}^2}{B},
\end{align}
where $(b)$ arises from the fact that for any \(k_1 \neq k_2\) and \(j_1 \neq j_2\), the conditional independence holds as
\begin{align}
\label{zero_eq}
&\mathbb{E}\biggl\{\left(\nabla F_{k_1}(\mathbf{w}_{k_1,t,j_1},\boldsymbol\xi_{k_1}^{j_1})- \nabla F(\mathbf{w}_{k_1,t,j_1})\right)^\top \times \nonumber\\
&\left(\nabla F_{k_2}(\mathbf{w}_{k_2,t,j_2},\boldsymbol\xi_{k_2}^{j_2})- \nabla F(\mathbf{w}_{k_2,t,j_2})\right)\biggr\} =\mathbb{E}_{{\xi}_{k_2}^{j_2}}\biggl\{ \mathbb{E}_{{\xi}_{k_1}^{j_1}}\biggl\{\nonumber\\
&\left(\nabla F_{k_1}(\mathbf{w}_{k_1,t,j_1},\boldsymbol\xi_{k_1}^{j_1})- \nabla F(\mathbf{w}_{k_1,t,j_1})\right)^\top\biggr\} \times \nonumber\\
&\left(\nabla F_{k_2}(\mathbf{w}_{k_2,t,j_2},\boldsymbol\xi_{k_2}^{j_2})- \nabla F(\mathbf{w}_{k_2,t,j_2})\right)|{\boldsymbol\xi}_{k_1}^{j_1}\biggr\} = 0,
\end{align}
where $\mathbb{E}_{{\xi}_{k_1}^{j_1}}\left\{\nabla F_{k_1}(\mathbf{w}_{k_1,j_1,t},\boldsymbol\xi_{k_1}^{j_1})- \nabla F(\mathbf{w}_{k_1,j_1,t})\right\} = \mathbf{0}$. Subsequently, $(c)$ is derived from Assumption 2. Substituting \eqref{RHS1} and \eqref{RHS2} into \eqref{normexp}, and then incorporating the resulting expression into \eqref{diffnorm}, we obtain
\begin{align}
\label{product}
&\mathbb{E}\left\{\nabla F( {\mathbf{w}}_{\text{G},t})^\top \nabla F(\mathbf{w}_{k,t,i})\right\} \geq \frac{1}{2} \mathbb{E}\left\{\|\nabla F( {\mathbf{w}}_{\text{G},t})\|^2\right\}\nonumber\\& + \frac{1}{2} \mathbb{E}\left\{\|\nabla F(\mathbf{w}_{k,i,t})\|^2\right\} - \frac{L^2 \mu_t^2}{2}i \sum_{j=0}^{i-1}\mathbb{E}\left\{\left\Vert\nabla F(\mathbf{w}_{k,t,j}) \right\Vert^2\right\}\nonumber\\&-\frac{L^2\mu_t^2}{2}i\frac{\sigma_\text{g}^2}{B}.
\end{align}
Consequently, we derive the subsequent bound
\begin{align}
\label{RHS1of}
&-\frac{\mu_{t}}{\mathbf{1}^\top\mathbf{a}_{t}}\sum_{k=1}^{K}a_{k,t} \sum_{i=0}^{\tau-1} \mathbb{E}\left\{\nabla F( {\mathbf{w}}_{\text{G},t})^\top \nabla F(\mathbf{w}_{k,t,i})\right\} \leq -\frac{\mu_{t}\tau}{2}\times\nonumber\\
& \mathbb{E}\left\{\|\nabla F( {\mathbf{w}}_{\text{G},t})\|^2\right\}-\frac{\mu_{t}}{2\mathbf{1}^\top\mathbf{a}_{t}}\sum_{k=1}^{K}a_{k,t} \sum_{i=0}^{\tau-1} \mathbb{E}\left\{\|\nabla F(\mathbf{w}_{k,i,t})\|^2\right\}\nonumber\\&+\frac{L^2\mu_{t}^3}{2\mathbf{1}^\top\mathbf{a}_{t}}\sum_{k=1}^{K}a_{k,t} \sum_{i=0}^{\tau-1} i \sum_{j=0}^{i-1}\mathbb{E}\left\{\left\Vert\nabla F(\mathbf{w}_{k,t,j}) \right\Vert^2\right\}\nonumber\\&+\frac{L^2\mu_{t}^3}{2}\frac{\tau (\tau-1)}{2}\frac{\sigma_\text{g}^2}{B}.
\end{align}
Subsequently, we provide an upper bound for the second term on the right side of \eqref{expected} as
\begin{align}
\label{weighted_norm}
&\mathbb{E}\left\{ \left\Vert \sum_{k=1}^{K}\frac{a_{k,t}}{\mathbf{1}^\top\mathbf{a}_{t}} \sum_{i=0}^{\tau-1}\nabla F_k(\mathbf{w}_{k,t,i},\boldsymbol\xi_{k}^i)\right\Vert^2\right\} \nonumber\\
&= \mathbb{E}\left\{ \left\Vert \sum_{k=1}^{K}\frac{a_{k,t}}{\mathbf{1}^\top\mathbf{a}_{t}} \sum_{i=0}^{\tau-1}\nabla F(\mathbf{w}_{k,t,i})\right\Vert^2\right\}+\nonumber\\
&\mathbb{E}\left\{ \left\Vert \sum_{k=1}^{K}\frac{a_{k,t}}{\mathbf{1}^\top\mathbf{a}_{t}} \sum_{i=0}^{\tau-1}\left(\nabla F_k(\mathbf{w}_{k,t,i},\boldsymbol\xi_{k}^i)-\nabla F(\mathbf{w}_{k,t,i})\right)\right\Vert^2\right\},
\end{align}
where
\begin{align}
\label{RRHS1}
&\mathbb{E}\left\{ \left\Vert \sum_{k=1}^{K}\frac{a_{k,t}}{\mathbf{1}^\top\mathbf{a}_{t}} \sum_{i=0}^{\tau-1}\nabla F(\mathbf{w}_{k,t,i})\right\Vert^2\right\} \nonumber\\&\stackrel{(d)}{\leq} \sum_{k=1}^{K}\frac{a_{k,t}}{\mathbf{1}^\top\mathbf{a}_{t}} \mathbb{E}\left\{ \left\Vert\sum_{i=0}^{\tau-1}\nabla F(\mathbf{w}_{k,t,i})\right\Vert^2\right\} \nonumber\\
&\stackrel{(e)}{\leq} \tau \sum_{k=1}^{K}\frac{a_{k,t}}{\mathbf{1}^\top\mathbf{a}_{t}} \sum_{i=0}^{\tau-1}\mathbb{E}\left\{ \left\Vert\nabla F(\mathbf{w}_{k,t,i})\right\Vert^2\right\}, 
\end{align}
where $(d)$ is a consequence of the convexity of $\|\cdot\|^2$ and $(e)$ stems from the arithmetic-geometric mean inequality. Also, the second term on the right side of \eqref{weighted_norm} can be upper-bounded as
\begin{align}
\label{RRHS2}
&\mathbb{E}\left\{ \left\Vert \sum_{k=1}^{K}\frac{a_{k,t}}{\mathbf{1}^\top\mathbf{a}_{t}} \sum_{i=0}^{\tau-1}\left(\nabla F_k(\mathbf{w}_{k,t,i},\boldsymbol\xi_{k}^i)-\nabla F(\mathbf{w}_{k,t,i})\right)\right\Vert^2\right\} =\nonumber
\end{align}
\begin{align}
& \sum_{k=1}^{K}\frac{a_{k,t}^2}{(\mathbf{1}^\top\mathbf{a}_{t})^2} \sum_{i=0}^{\tau-1}\mathbb{E}\left\{ \left\Vert\left(\nabla F_k(\mathbf{w}_{k,t,i},\boldsymbol\xi_{k}^i)-\nabla F(\mathbf{w}_{k,t,i})\right)\right\Vert^2\right\} \nonumber\\
& \leq \frac{\sigma_\text{g}^2}{B}\tau \sum_{k=1}^{K}\frac{a_{k,t}^2}{(\mathbf{1}^\top\mathbf{a}_{t})^2},
\end{align}
derived from \eqref{zero_eq} and Assumption 2. Replacing \eqref{RRHS1} and \eqref{RRHS2} in \eqref{weighted_norm} and replacing the result with \eqref{RHS1of} in \eqref{expected}, we obtain
\begin{align}
&\mathbb{E}\left\{F( {\mathbf{w}}_{\text{G},t+1}) - F( {\mathbf{w}}_{\text{G},t}) \right\} \leq  -\frac{\mu_{t}\tau}{2} \mathbb{E}\left\{\|\nabla F( {\mathbf{w}}_{\text{G},t})\|^2\right\}\nonumber\\&-\frac{\mu_{t}}{2\mathbf{1}^\top\mathbf{a}_{t}}\sum_{k=1}^{K}a_{k,t} \sum_{i=0}^{\tau-1} \mathbb{E}\left\{\|\nabla F(\mathbf{w}_{k,i,t})\|^2\right\}+\frac{L^2\mu_{t}^3}{2\mathbf{1}^\top\mathbf{a}_{t}}\sum_{k=1}^{K}a_{k,t}\nonumber\\
& \sum_{i=0}^{\tau-1} i \sum_{j=0}^{i-1}\mathbb{E}\left\{\left\Vert\nabla F(\mathbf{w}_{k,t,j}) \right\Vert^2\right\}+\frac{L^2\mu_{t}^3}{2}\frac{\tau (\tau-1)}{2}\frac{\sigma_\text{g}^2}{B}+\frac{L\mu_t^2}{2}\tau \nonumber\\
& \sum_{k=1}^{K}\frac{a_{k,t}}{\mathbf{1}^\top\mathbf{a}_{t}} \sum_{i=0}^{\tau-1}\mathbb{E}\left\{ \left\Vert\nabla F(\mathbf{w}_{k,t,i})\right\Vert^2\right\}+\frac{L \mu_t^2}{2}\frac{\sigma_\text{g}^2}{B}\tau \sum_{k=1}^{K}\frac{a_{k,t}^2}{(\mathbf{1}^\top\mathbf{a}_{t})^2} \nonumber\\& +\frac{L}{2} {\text{QMSE}_t(\mathbf{a}_{t})}, 
\end{align}
where, by leveraging the bound 
\begin{align}
&\sum_{i=0}^{\tau-1} i \sum_{j=0}^{i-1} \mathbb{E}\left\{\left\Vert\nabla F(\mathbf{w}_{k,t,j}) \right\Vert^2\right\} \nonumber\\
&\leq \sum_{i=0}^{\tau-1} i \times \sum_{i=0}^{\tau-1} \mathbb{E}\left\{\left\Vert\nabla F(\mathbf{w}_{k,t,i}) \right\Vert^2\right\} \nonumber\\
&= \frac{\tau(\tau-1)}{2}\sum_{i=0}^{\tau-1} \mathbb{E}\left\{\left\Vert\nabla F(\mathbf{w}_{k,t,i}) \right\Vert^2\right\},
\end{align}
we obtain the following bound on \eqref{expected}
\begin{align}
\label{mid_stepppp}
&\mathbb{E}\left\{F( {\mathbf{w}}_{\text{G},t+1}) - F( {\mathbf{w}}_{\text{G},t}) \right\} \leq  -\frac{\mu_{t}\tau}{2} \mathbb{E}\left\{\|\nabla F( {\mathbf{w}}_{\text{G},t})\|^2\right\}-\nonumber\\
&\frac{\mu_{t}}{2\mathbf{1}^\top\mathbf{a}_{t}}\sum_{k=1}^{K}a_{k,t} \sum_{i=0}^{\tau-1} \mathbb{E}\left\{\|\nabla F(\mathbf{w}_{k,i,t})\|^2\right\}+\frac{L^2\mu_{t}^3}{2\mathbf{1}^\top\mathbf{a}_{t}}\frac{\tau(\tau-1)}{2}\nonumber\\
&\sum_{k=1}^{K}a_{k,t} \sum_{i=0}^{\tau-1} \mathbb{E}\left\{\left\Vert\nabla F(\mathbf{w}_{k,t,i}) \right\Vert^2\right\}+\frac{L^2\mu_{t}^3}{2}\frac{\tau (\tau-1)}{2}\frac{\sigma_\text{g}^2}{B}+\nonumber\\
&\frac{L\mu_t^2}{2}\tau \sum_{k=1}^{K}\frac{a_{k,t}}{\mathbf{1}^\top\mathbf{a}_{t}} \sum_{i=0}^{\tau-1}\mathbb{E}\left\{ \left\Vert\nabla F(\mathbf{w}_{k,t,i})\right\Vert^2\right\}+\frac{L \mu_t^2}{2}\frac{\sigma_\text{g}^2}{B}\tau \nonumber\\
& \sum_{k=1}^{K}\frac{a_{k,t}^2}{(\mathbf{1}^\top\mathbf{a}_{t})^2}  +\frac{L}{2} {\text{QMSE}_t(\mathbf{a}_{t})} \nonumber\\&=  -\frac{\mu_t}{2}\left(1-\frac{L^2\mu_t^2}{2}\tau(\tau-1)-L\mu_t \tau\right)\sum_{k=1}^{K}\frac{a_{k,t}}{\mathbf{1}^\top\mathbf{a}_{t}} \nonumber\\
&\sum_{i=0}^{\tau-1} \mathbb{E}\left\{\|\nabla F(\mathbf{w}_{k,i,t})\|^2\right\}+\frac{L^2\mu_{t}^3}{2}\frac{\tau (\tau-1)}{2}\frac{\sigma_\text{g}^2}{B}+\frac{L \mu_t^2}{2}\frac{\sigma_\text{g}^2}{B}\tau \nonumber\\& \sum_{k=1}^{K}\frac{a_{k,t}^2}{(\mathbf{1}^\top\mathbf{a}_{t})^2} +\frac{L}{2} {\text{QMSE}_t(\mathbf{a}_{t})}-\frac{\mu_{t}\tau}{2} \mathbb{E}\left\{\|\nabla F( {\mathbf{w}}_{\text{G},t})\|^2\right\}.
\end{align}
Given the condition
\begin{align}
1-\frac{L^2\mu_t^2}{2}\tau(\tau-1)-L\mu_t \tau \geq 0,
\end{align}
we have
\begin{align}
&-\frac{\mu_t}{2}\left(1-\frac{L^2\mu_t^2}{2}\tau(\tau-1)-L\mu_t \tau\right)\times\nonumber\\&\sum_{k=1}^{K}\frac{a_{k,t}}{\mathbf{1}^\top\mathbf{a}_{t}}
\sum_{i=0}^{\tau-1} \mathbb{E}\left\{\|\nabla F(\mathbf{w}_{k,i,t})\|^2\right\} \leq 0,
\end{align}
and hence, we can bound \eqref{mid_stepppp} as
\begin{align}
&\mathbb{E}\left\{F( {\mathbf{w}}_{\text{G},t+1}) - F( {\mathbf{w}}_{\text{G},t}) \right\} \leq  -\frac{\mu_{t}\tau}{2} \mathbb{E}\left\{\|\nabla F( {\mathbf{w}}_{\text{G},t})\|^2\right\}+\nonumber\\&\frac{L^2\mu_{t}^3}{2}\frac{\tau (\tau-1)}{2}\frac{\sigma_\text{g}^2}{B}+\frac{L \mu_t^2}{2}\frac{\sigma_\text{g}^2}{B}\tau \sum_{k=1}^{K}\frac{a_{k,t}^2}{(\mathbf{1}^\top\mathbf{a}_{t})^2} +\frac{L}{2} {\text{QMSE}_t(\mathbf{a}_{t})}.
\end{align}
After applying Assumption 3, we have for any $t \in \left\{0,\ldots,T-1\right\}$
\begin{align}
\label{finalstep}
&\mathbb{E}\left\{F({\mathbf{w}}_{\text{G},t+1})\right\}-F^* \leq {\left(1-\mu_t \tau \delta\right)} \biggl(\mathbb{E}\left\{F({\mathbf{w}}_{\text{G},t})\right\}-F^*\biggr)+\nonumber\\&\frac{L^2\mu_{t}^3}{2}\frac{\tau (\tau-1)}{2}\frac{\sigma_\text{g}^2}{B}+\frac{L \mu_t^2}{2}\frac{\sigma_\text{g}^2}{B}\tau \sum_{k=1}^{K}\frac{a_{k,t}^2}{(\mathbf{1}^\top\mathbf{a}_{t})^2} +\frac{L}{2} {\text{QMSE}_t(\mathbf{a}_{t})}.
\end{align}
This bound links the steps $t+1$ and $t$. To determine the bound specified in Theorem 3, we can substitute 
\( \mathbb{E}\left\{ F({\mathbf{w}}_{\text{G},t}) \right\} - F^* \)
on the right side with the equivalent one-step bound for the steps \( t \) and \( t-1 \). By consistently applying this procedure over the interval 
\( \left\{ t-1,\ldots,0 \right\} \),
the proof is complete.

\end{document}